\documentclass{aa}  
\usepackage{graphicx}
\usepackage{txfonts}
\usepackage{subcaption}
\usepackage{verbatim}

\usepackage[switch]{lineno}
\usepackage{datetime}
\usepackage[usenames,table,dvipsnames]{xcolor}
\usepackage{setspace}
\usepackage[normalem]{ulem}
\usepackage{soul}
\usepackage{hyperref}
\usepackage{multirow}

\definecolor{blue}{RGB}{66, 153, 233}
\definecolor{red}{RGB}{255, 0, 0}
\definecolor{purple}{RGB}{255, 0, 255}

\newcount\Comments  
\newcommand{\kibitz}[2]{\ifnum\Comments=1\textcolor{#1}{#2}\fi}

\begin{document}

   \title{Bayesian insights in Tycho supernova remnant : a detailed mapping of ejecta properties}


   \author{L. Godinaud\inst{1}
          \and
          F. Acero \inst{2,3}
          \and 
          A. Decourchelle \inst{2}
          \and 
          J. Ballet \inst{2}
          }

   \institute{Université Paris Cité, Université Paris-Saclay, CEA, CNRS, AIM, F-91191, Gif-sur-Yvette, France \\
              \email{leila.godinaud@cea.fr}
              \and
            Université Paris-Saclay, Université Paris Cité, CEA, CNRS, AIM, 91191, Gif-sur-Yvette, France \\
              \email{fabio.acero@cea.fr}
              \and
            FSLAC IRL 2009, CNRS/IAC, La Laguna, Tenerife, Spain \\              
             }


  \abstract
    {While Tycho's supernova remnant is one of the most studied type Ia Galactic supernova remnants, a global view of the physical properties of its ejecta is lacking, to understand its mysteries. In particular, the spatial distribution of the Si-rich ejecta line-of-sight velocity presents a large-scale unexplained asymmetry, with the north dominantly blueshifted and the south redshifted.}
   {To investigate the origin of this line-of-sight velocity asymmetry in the ejecta  and its current dynamics, we carry out a detailed X-ray spatially-resolved spectral analysis of the entire shocked ejecta in Tycho's SNR to determine the physical properties of its various components. This study is based on the archival deep X-ray observations from the Chandra space telescope.} 
   {The spatially-resolved spectral analysis in 211 regions over the entire SNR is based on a tesselation method applied to the line-of-sight velocity map. We model the ejecta emission with two thermal non-equilibrium ionisation components of different compositions for intermediate-mass elements (IME) and iron-rich ejecta. We include Doppler shift and line broadening and add a power law for the synchrotron emission, and additional constraints.  A Bayesian tool is used to conduct the fitting, using a nested sampling algorithm. It allows us to obtain a complete view of the statistical landscape. }
   {We provide maps of the physical parameters of the various components across the SNR ejecta. The Doppler shift map confirms spectrally the large-scale north-south asymmetry in the line-of-sight velocity that was obtained from a general morphological component analysis. We reveal different spatial distributions of temperature and ionization time for IMEs and for iron-rich ejecta, but none of these maps shows structure associated to the large-scale north-south asymmetry in the line-of-sight velocity distribution.
   In the IME component, we observe an overall anti-correlation between the temperature and ionization time which could arise from different ionization histories.
   The abundance maps show spatial variations, depending on the element, perhaps due to an origin in different layers during the explosion. We compare these abundances with some nucleosynthesis models. In addition, we observe for the first time an emission line at 0.654 keV possibly related to oxygen. Its spatial distribution differs from the other elements, so that this line may arise in the ambient medium. }
   {}


   \maketitle
%

\section{Introduction}

In the last 500 yrs, only two Galactic supernovae were witnessed by eye on Earth. They occured on November 1572 (Tycho's supernova ; SN 1572) and on October 1604 (Kepler's supernova ; SN 1064). Both supernovae are known to be associated with the type Ia thermonuclear supernova class but the exact mechanism leading to the explosion is still unclear \citep[see e.g.][for a review]{Liu2023}.
In the single degenerate scenario (SD scenario), a degenerate carbon-oxygen white dwarf accretes matter from a stellar companion. During its life, this companion produces winds that will create a dense circumstellar medium (CSM) in which the remnant of the supernova will evolve. After the explosion, this non-degenerate second star is expected to survive and be observable centuries after.
The alternative scenario is the double degenerate (DD scenario) where the progenitor system consists of two degenerate white dwarfs. In this case, the CSM is less dense and there is no surviving donor, in contrast to the SD channel.
It is expected that the SD or DD scenarios will leave different  fingerprints on the supernova remnant (SNR) that we can now study, centuries after the explosion. For example, the different explosion channels will lead to contrasting nucleosynthesis yields, impacting the ejecta abundances that can be measured using X-ray observations of the remnant. 
The explosion mechanism can also produce asymmetries in the ejecta velocity distribution. Note that the velocity asymmetry that we observe now can also be acquired due to interaction with an inhomogeneous CSM or interstellar medium (ISM).

In the case of Tycho's SNR, the precise type Ia scenario of explosion has not been clarified yet. 
X-ray observations of Tycho's SNR provide constraints on the abundance and spatial distribution of elements in the ejecta, and revealed significant abundances variations, in particular in the southeastern knots \citep[e.g.][]{Decourchelle2001, Williams2020, Yamaguchi2017}. By modeling the global X-ray spectrum, \cite{Badenes2006} favored a “normal” SN Ia event, which was later confirmed by optical spectroscopy of light echoes from SN1572 \citep{Krause2008}.  
Its ambient medium is not very dense but remains complex. A gradient of external density from west to east is derived from infrared observations \citep{Williams2013}.
A SD scenario is suggested but there is still no consensus since no surviving donor has been identified \citep{Kerzendorf2018, Godinaud2023}. 

Studies of the dynamical properties of the ejecta were performed in Tycho's SNR and highlighted azimuthal variations of the velocity in the plane of the sky. A fast iron-rich knot has been studied by \cite{Yamaguchi2017}, and is interpreted as a dense ejecta bullet produced during the explosion. A slowdown in the north-eastern rim was revealed by \cite{Godinaud2023}, which is probably due to an interaction with a potential molecular cloud observed in radio \citep{Zhou2016}. A gradient of external density is also observed \citep{Williams2017} from west to east, which is correlated with the dynamics.
For the velocity in the line of sight (hereafter $V_{\rm z}$), a large-scale asymmetry was also observed, with the ejecta in the north being blueshifted and in the south dominantly redshifted \citep{Millard2022, Godinaud2023, Uchida2024}. The origin of this asymmetry is not yet understood. It can be innate, due to the explosion, or acquired after the interaction with the CSM or interstellar clouds. In the latter case, these interactions are expected to impact the ejecta physical parameters, such as the temperature or the ionization time. 
To understand this large-scale asymmetry in the line-of-sight velocity distribution of Si-rich ejecta, and investigate the origin of Tycho's SNR, we perform a detailed spatially-resolved spectroscopic analysis of the whole ejecta material in the X-ray band.

In the last two decades, the deep observations ($>$ 100 ks) of the brightest SNRs have provided enough photon counts to obtain detailed physical parameter maps of entire young SNRs through spatially-resolved spectroscopy. This work has been carried out for example to study the ejecta properties of type Ia SNR candidates. From the mapping of Kepler's SNR, \cite{Sun2019} highlighted the properties of the ejecta and ambient medium, in particular their abundances. The properties of the thermal and non-thermal components of the SN 1006 remnant were measured by \cite{Li2015}. The mapping of the W49B SNR by \cite{Zhou2018}, indicates a gradient in physical properties, therefore an inhomogeneous environment.

The goal of this study is to perform a detailed parameter mapping of the ejecta properties through a complete spatially-resolved spectral study to better understand the observed asymmetries and abundances in order to shed light on the explosion mechanisms.
In parallel to this work, \cite{Uchida2024} performed a parameter mapping of Tycho's SNR with XMM-Newton data. They mostly focused on the interpretation of $V_{\rm z}$ ejecta dynamics on the rim of the SNR. By using a hydrodynamical simulation, they concluded that Tycho's SNR is likely interacting with a dense circumstellar wall, which supports a SD origin.

To perform a mapping of the SNR parameters, a segmentation of the extended source into a large number of spectral extraction regions is the first step. As presented in Sect \ref{Section:Segmentation and spectrum extraction}, we chose to base our tessellation on the $V_{\rm z}$ map derived in \citet{Godinaud2023} to have regions homogeneous in velocity.
The different steps and choices leading to our spectral model are presented in Sect \ref{Section:Model et fitting}.
Given the large dimensionality of such a model and possible resulting degeneracies, traditional fitting methods based on gradient descent lack robustness and we ended up using a nested sampling algorithm to fit our high dimension spectral model. We used the Bayesian X-ray analysis framework \citep[hereafter BXA, see][]{Buchner2014}. Though it has recently been applied in a study of the Tycho's SNR by \cite{Ellien2023} to study the surrounding medium in a few regions, this is the first time that a nested sampling algorithm is used at large scale for parameter mapping in a SNR.
We describe the results of our parameter maps in Sect \ref{Section:Results}. The origin of the $V_{\rm z}$ asymmetry is discussed in Sect \ref{Section:Discussion} as well as a selection of scientific results obtained for the first time with this mapping. Finally Sect \ref{Section:Conclusion} summarizes this study.


\section{Segmentation and spectral extraction}
\label{Section:Segmentation and spectrum extraction}

\begin{figure*}
\centering
\includegraphics[scale=0.4, trim = 0 0 0 0, clip=true]{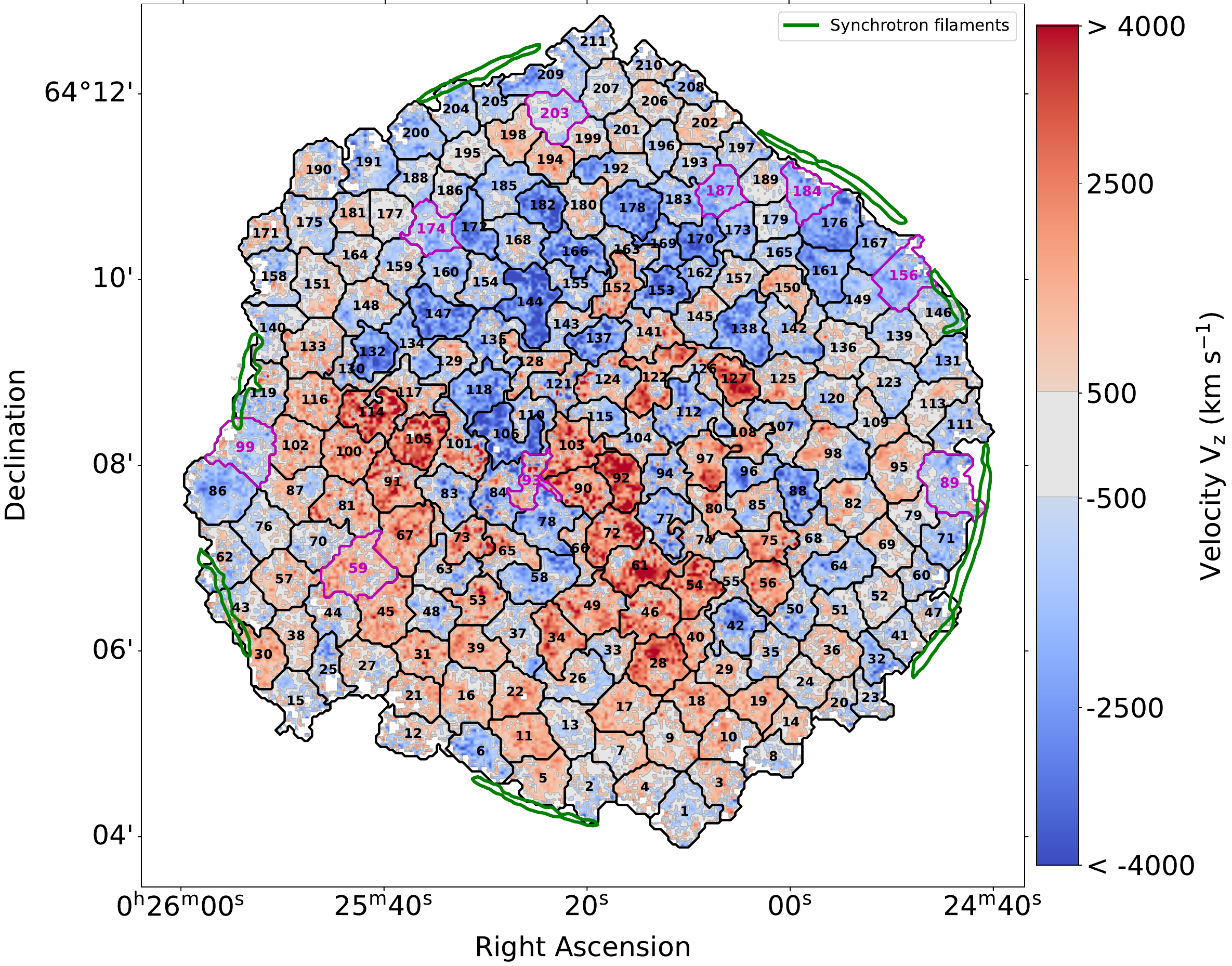}
\caption{\footnotesize Map of the integrated Si line velocity in the line of sight $V_{\rm z}$ \citep[obtained in][]{Godinaud2023} and the segmentation of the SNR for spectral extraction used in this study. The segmentation is optimized to achieve $V_{\rm z}$ homogeneity in each region. The region number is a convention which goes from south to north.The forward shock, seen in X-ray synchrotron emission, is shown in green contours.}
\label{fig:Segmentation}
\end{figure*}

The first step of our study is to define a method for the segmentation of the whole SNR into small regions from which the spectra will be extracted. The typical method used is adaptive binning as the Voronoi \citep{diehl06b} or contour \citep{sanders06} binning algorithm to adapt the size of the region to reach an equivalent number of photons in each region. Nevertheless, the flux map is not necessarily indicative of the underlying physical structures of the SNR. These structures are further complicated by the integration along the line of sight that mixes the front and rear of the SNR shell.

As one of the motivations of this work is to explore whether the large-scale north/south line-of-sight velocity asymmetry is linked to other physical properties of the ejecta plasma, we choose to base our segmentation on the map of integrated velocity in the line of sight $V_{\rm z}$ obtained in \cite{Godinaud2023}. Figure \ref{fig:Segmentation} presents this map and the segmentation for the extraction regions, showing zones dominated at small scale by one side or another of the SNR.

In practice, we use the \emph{Slic} algorithm from the \emph{scikit-image} python package \footnote{\href{https://scikit-image.org/docs/stable/auto_examples/segmentation/plot_segmentations.html}
{Tutorial on the \emph{Skimage} website.}}, which has two main parameters: the number of regions and a compactness factor. 
The number of regions is a compromise between two elements. With a high number of regions, the map will be more spatially resolved and each region will be more homogeneous in terms of physical parameters. However the spectrum from each region will have lower statistics reducing our physical constraints on each parameter. For the compactness, we aim for a good homogeneity of $V_{\rm z}$ but each region must also be compact enough to correspond to the same spatial structure. We found a good compromise with compactness factor at 1000 and with 211 regions as shown in Figure \ref{fig:Segmentation}. 

With this segmentation, the regions correspond to homogeneous line-of-sight velocities, which will help the spectral modeling. Note that the $V_{\rm z}$  map of \citet{Godinaud2023} is restricted to Si-emitting ejecta regions and excludes the thin regions along the rim lying between the outermost ejecta and the forward shock.
To place this region definition in a broader context, the segmentation is overlaid on the flux image in two different energy bands in App \ref{Appendix : Segmentation}.

Then, we extract the X-ray spectra from \emph{Chandra}'s archival data of the nine deep observations\footnote{ObsId in the \emph{Chandra}'s archive : 10093, 10094, 10095, 10096, 10097, 10902, 10903, 10904, 10906.} from 2009 (734 ks in total) as summarized in \cite{Godinaud2023}. The tool \emph{specextract} is used to obtain for each region and each observation, the spectra and responses files.
The median value of the region surface is 900 arcsec$^2$. In the energy band [0.5, 7] keV, there are in average around 2$\times$10$^5$ photons in each region. The minimum is around 4$\times$10$^4$ photons in region 93, which is located at the center of the CCD gaps of Chandra ACIS-I.
The extraction procedure results in 211$\times$9 (1899) spectra. For each region a joint fit of all observations is performed.

\section{Spectral modeling and parameter fitting}
\label{Section:Model et fitting}

The global X-ray spectrum of Tycho's SNR shows prominent K-shell line emission of Si, S, Ar, Ca and Fe, as well as Fe L-shell line emission \citep{Decourchelle2001}. Non-thermal emission is associated with the forward shock \citep{Hwang2002}, whose synchrotron emission arises from TeV accelerated electrons \citep{Ballet2006}. The thermal emission is dominated by the shocked ejecta. The emission from the shocked ambient medium is expected to be much fainter and has proven to be difficult to detect even by detailed X-ray spectral studies of selected post-shock regions along the shock wave \citep{Cassam-Chenai2007}. Constraints have been obtained recently by \cite{Ellien2023}, who estimated a mean temperature of ~1 keV and a density of $\sim$0.3 cm$^{-3}$ for this component. As previously explained, these thin regions along the shock (see Fig. \ref{fig:Segmentation}) are by construction excluded from our study of the ejecta properties.

Constructing a model that is sufficiently detailed to account for all the plasma components and observed spectral features in spectra with such a high statistics, and at the same time general enough to cope with the  diversity of behaviors observed across the SNR, is a challenging task.
This section first presents the different steps that led us to our final spectral model, then introduces the necessary parameter constraints, and then the method used to optimize all our parameters.

\subsection{Construction of the model}
\label{Subsection:Model}

In the X-ray spectrum of Tycho's SNR, we expect the contributions of at least three different physical entities: the shocked ejecta, the high-energy shock-accelerated electrons, and the shocked ambient medium. The last is deemed to be negligible. The shocked ejecta contribution potentially contains various components, associated to regions of different supernova composition and history. The intermediate mass elements (IME) and the iron-rich ejecta are known to have different conditions, with iron at a higher temperature and lower ionization than Si \citep{Hwang1997}.

In our model, we will thus consider two non-equilibrium ionization collisional plasma models with variable abundances, one for the IME ejecta (VNEI1) and one for the iron group ejecta (VNEI2). To take into account the ejecta dynamics in the line of sight, we use the redshift parameter. In addition, we include a power-law component to account for the synchrotron emission of the TeV accelerated electrons, whose emission is dominant between 4 and 6 keV. All these components undergo a certain level of absorption $N_{\rm H}$ at low energy due to the intervening interstellar material in the line of sight, modeled by TBABS. We perform the spectral analysis over the energy band between 0.5 and 7 keV.

The first objective of this paper is to determine the Doppler shift map of the Si line from a spatially-resolved spectral analysis of Tycho's SNR and to compare the results to the map obtained with a general morphological component analysis \citep{Godinaud2023}.
The second objective is to study the properties of the IMEs to investigate the origin of the observed $V_{\rm z}$ asymmetry in the Si-line Doppler shift map shown in Fig \ref{fig:Segmentation}. This asymmetry can be innate due to the explosion or acquired during the expansion in a circumstellar or inhomogenous ambient medium. By mapping the properties of the ejecta, and specifically those of the Si element, we search for potential signatures associated with this asymmetry and for traces of large scale interactions with the surrounding medium.

In this first attempt to model the 211 spectra, some important spectral residual structures were observed with some wings around the silicon line (the observed Si line being larger than what our model predicts). These residuals were observed to be more prominent at the center of the SNR.
We interpreted this line broadening due to the opposite dynamics of the two half-shells in the line of sight (two Si lines ; a blue and redshifted one). An additional broadening in each shell caused by variations of plasma properties may also play a role.
As it is difficult to disentangle these components at Chandra's spectral resolution, we chose to account for this effect with a convolutional Gaussian smoothing on the ejecta components using Xspec's \textit{gsmooth} model with the parameter \textit{Sig$_{\rm 6keV}$} free to vary as it has been done in previous studies \citep{Kasuga2021, Ellien2023}.

Additional excesses in the residuals are observed at 1.24 keV and 0.72 keV in most spectra. In \cite{Yamaguchi2011}, this was interpreted as iron lines missing from the atomic model. We thus add a Gaussian centered at 1.24 keV (hereafter \emph{G1}) and another at 0.72 keV (hereafter \emph{G2}).
At low energy, some residuals were also observed around 0.57 and 0.65 keV possibly due to oxygen emission lines.
This oxygen emission could originate either from the ejecta, but perhaps with different properties than the IME thermal component, or from the shocked ambient medium that should be modeled with another NEI component.
But there is not enough information in this low energy band to introduce and constrain such an additional physical component.
So we add two Gaussian components at 0.654 keV (hereafter \emph{G3}) and 0.57 keV (hereafter \emph{G4}) where oxygen lines are expected if this element emits.
For all four gaussian lines, their sigma is equal to zero and their redshift is linked to the ejecta using the \textit{zgauss} model. Only their normalizations are allowed to vary.
 
 The final model, presented in Table \ref{table:Model}, has 19 free parameters and is summarized by :
\( Tbabs(gsmooth(VNEI\ 1 + VNEI\ 2 + zgauss\ (\ G1 +\ G2 + \ G3 + \ G4)) + powerlaw) \).

The abundance table that was used is \emph{wilm} \citep{Wilms2000}.

\begin{table*}[h!]
\caption{\footnotesize Xspec model used in this study. It consists of two thermal non-equilibrium ionization collisional components and a power law. The first represents the IMEs (VNEI 1) and the latter the iron-rich plasma (VNEI 2). Four Gaussians are added to account for line features in the residuals. Their sigma is frozen to 0 and the redshift is linked to that of the ejecta VNEI (z1). 
The status of the parameter during the fit is indicated and its value if it is frozen. If the parameter is free to vary, we indicate the type of prior : logarithmic uniform ($\mathcal{LU}$), linearly uniform ($\mathcal{U}$) or Gaussian ($\mathcal{G}$). The last two columns are the median and standard deviation (STD) of each parameter over all regions.}
\label{table:Model}    
\centering                         
\begin{tabular}{|c|c c|c|c|c||c c|}        
\hline                 
Component & parameter & units & status & value & prior  & median & STD \\ 
\hline \hline                       
\multirow{1}{5em}{TBabs} & $N_{\rm H}$ & $10^{22}$  cm$^{-2}$ & thawed & - &  $\mathcal{G}$(0.6, 0.1) & 0.70 & 0.06 \\ 
\hline
\multirow{2}{5em}{gsmooth} & Sig$_{\rm 6keV}$ & keV & thawed & - & $\mathcal{LU}$(6e-3, 0.15) & 0.08 & 0.02 \\ 
& Index &  & frozen & 1 &  - & - & - \\ 
\hline
\multirow{12}{5em}{VNEI 1} & k$T$ & keV & thawed & - & $\mathcal{LU}$(0.4, 10)  & 1.3 & 0.7 \\ 
& H, He, C, N, O, Ne &  & frozen & 1 & - & - & - \\ 
& Mg &  & thawed & - & $\mathcal{LU}$(0.1, 500) & 49 & 13 \\ 
& Si &  & frozen &  1000 & -  & - & - \\ 
& S &  & thawed & - & $\mathcal{LU}$(50, 9999) & 762 & 82 \\ 
& Ar &  & thawed & - & $\mathcal{LU}$(50, 9999) & 772 & 182 \\ 
& Ca &  & thawed & - & $\mathcal{LU}$(50, 9999) & 1754 & 690 \\ 
& Fe, Ni &  & frozen &  1 & -  & - & - \\ 
& Tau & cm$^{-3}$ s & thawed & - & $\mathcal{LU}$(5e9, 5e12) & 4.7e10 & 1.3e10 \\ 
& Redshift z1 &  & thawed & - & $\mathcal{U}$(-6e-2, 6e-2) & 0 & 5e-3 \\ 
& normalization & cm$^{-5}$ & thawed & - & $\mathcal{LU}$(1e-7, 1e-3) &  & \\ 
\hline
\multirow{8}{5em}{VNEI 2} & k$T$ & keV & thawed & - & $\mathcal{G}$(10, 2) & 9.0 & 1.5 \\ 
& H, He, C, N &  & frozen & 1 & -  & - & - \\ 
& O, Ne, Mg, Si, S, Ar, Ca &  & frozen &  1 & -  & - & - \\ 
& Fe &  & frozen & 1000 & -  & - & - \\ 
& Ni&  & frozen & 1000 & -  & - & - \\ 
& Tau & cm$^{-3}$ s & thawed & - & $\mathcal{LU}$(5e8, 5e11) & 9.0e9 & 1.9e9 \\ 
& Redshift z2 &  & linked to z1 & - & - & - & - \\ 
& normalization &  cm$^{-5}$ & thawed & - & $\mathcal{LU}$(1e-8, 1e-4) &  & \\ 
\hline
\multirow{2}{5em}{zgauss G1} & LineE & keV & frozen & 1.24 & -  & - & - \\ 
& normalization &  ph cm$^{-2}$s$^{-1}$ & thawed & - & $\mathcal{LU}$(1e-7, 1e-3) &  &  \\
\hline
\multirow{2}{5em}{zgauss G2} & LineE & keV & frozen & 0.72 & -  & - & - \\ 
& normalization & ph cm$^{-2}$s$^{-1}$   & thawed & - & $\mathcal{LU}$(1e-8, 1e-3) &  &  \\
\hline
\multirow{2}{5em}{zgauss G3} & LineE & keV & frozen & 0.654 & -  & - & - \\ 
& normalization &  ph cm$^{-2}$s$^{-1}$  & thawed & - & $\mathcal{LU}$(1e-8, 1e-3) &  &  \\
\hline
\multirow{2}{5em}{zgauss G4} & LineE & keV & frozen & 0.570 & -  & - & - \\ 
& normalization &  ph cm$^{-2}$s$^{-1}$ & thawed & - & $\mathcal{LU}$(1e-8, 1e-3) &  &  \\
\hline
\multirow{2}{5em}{power law} & PhoIndex &  & thawed & - & $\mathcal{G}$(2.6, 0.5) & 2.5 &  0.2 \\ 
& normalization at 1 keV &  ph cm$^{-2}$s$^{-1}$keV$^{-1}$ & thawed & - & $\mathcal{LU}$(1e-7, 2e-3) &  & \\ 
\hline

\hline                                   
\end{tabular}
\end{table*}

\subsection{Additional parameter constraints}
\label{Subsection:Parameters constraints}

At this level of statistics, the combined spectra in each region are very rich in spectral information even up to 7 keV, nevertheless, some additional constraints (shown in Table \ref{table:Model}) are necessary.

For the ejecta (negligible hydrogen), there is an important degeneracy between abundances and the normalization of the thermal components. To solve it, we fixed the abundance of the silicon for VNEI 1 to a value of 1000, as well as that of the iron for VNEI 2. The other abundances in the ejecta component are interpreted with respect to Si and Fe, respectively. The values are set to 1000 to represent a plasma in the pure ejecta regime as shown in \cite{Greco2020}.
We tested to fit the abundance of O and Ne, but they were not constrained in most of the regions. We thus fixed H, He, C, N, O, Ne, Fe and Ni to 1 in VNEI 1; H, He, C, N, O, Ne, Mg, Si, S, Ar, Ca to 1 in VNEI 2. This amounts to neglecting them.

As the redshift of the hot Fe component is not well constrained, we linked it to the redshift of the Si component. We suppose that the two components have the same dynamics, which is not always true (for example for the known iron knot in the South-East).

As all the components are not necessarily present in each of the 211 regions, several parameters are found to be difficult to constrain in some regions and can introduce degeneracies.
This is in particular the case for the synchrotron component, for which the spectral index cannot be constrained in regions where non-thermal emission is nearly absent. 
Similarly, when the Fe-K emission line is not observed, the temperature of the iron VNEI component is poorly constrained using only the Fe-L information.
To mitigate this effect, we introduce Gaussian priors for some specific parameters in the Bayesian fit (presented in the next section).
The absorption parameter $N_{\rm H}$  and the photon index of the power law are relatively uniform \citep{Williams2017}, so we use Gaussian priors centered respectively on $0.6 \times 10^{22}$ cm$^{-2}$ with a sigma of 0.1 $\times$ 10$^{22}$ cm$^{-2}$ for $N_{\rm H}$ and on 2.6 with a sigma of 0.5 for the photon index of the power law.
A similar approach was used in \cite{Mayer2023} to constrain the power law index.
We add also a Gaussian prior on the temperature of the iron VNEI to help the fit in low Fe-K line emission regions. Indeed, if there is very faint emission in the Fe-K line, which happens near the border of the SNR, the iron temperature is artificially biased towards lower values and not well constrained, resulting in biases on the ionization time as well. So we used the mean k$T_2$ values found in the rest of the SNR to set a Gaussian prior centered on 10 keV with a sigma of 2 keV.

\subsection{Parameter estimation with Nested Sampling}
\label{Subsection:BXA}

The model described in the previous section, which is designed to reproduce the diversity of spectral components in Tycho's SNR, has a high number of dimensions (19 degrees of freedom; d.o.f) and is thus very challenging to optimize. 
In addition to the dimensionality of the model, we also face the challenge of analysing a large number of spectra with 211 regions, each with a set of nine observations. 
While a model with a high number of dimensions can be optimized manually with human intervention (freezing/thawing some parameters, investigating the residuals and iterating), this procedure is  not suitable for analysing a large number of regions.

We first attempted to perform a spectral fit of each region with Xspec, but we realized that the convergence of the models was not robust. In other words, for a given spectrum, different starting points resulted in very different best fits.
The Xspec \verb|error| and \verb|steppar| commands were used to try to \textit{shake} the fit out of its local minimum without success.
Xspec's different fitting methods  Levenberg-Marquardt, Minuit2 (migrad method), or Minuit2 (simplex method) all showed similar issues.
We interpret this issue as the minimizer falling in a local minimum and not being able to find the true minimum. We note that this issue is particularly critical in our case due to the following challenges : 1) the high dimensionality of our problems (19 d.o.f.) , 2) due to the very high statistics in each region ($\sim 10^{5}$ counts), the likelihood minimum peak is very sharp and can easily be missed, and 3) several parameters (such as k$T$ and $\tau$, see Sect. \ref{subsection:Relation kT tau}) are strongly correlated.

To overcome these issues, we turned to the nested sampling (NS) algorithm, which is a Monte Carlo algorithm for computing an integral of the likelihood function over the model parameter space. The method performs this integral by evolving a collection of points through the parameter space \citep[see][for a recent review]{Ashton2022,Buchner2023}.

Without going into too many details, one important specificity of the NS method is that it starts from the entire parameter space and evolves a collection of live points to map all minima (including multiple modes if any) whereas Markov Chain Monte Carlo methods require an initialization point and the walkers will explore the local likelihood. The ability of these walkers to escape a local minimum or to accurately describe a complex likelihood space is not guaranteed.

To apply the NS method to our X-ray spectral analysis, we used the Bayesian X-ray analysis framework \citep[BXA v4.1.1,][]{Buchner2014}, which connects the nested sampling algorithm UltraNest \citep[v3.6.3,][]{Buchner2021} with the X-ray fitting environment CIAO/Sherpa \citep{Fruscione2006}.
The spectral analysis using this framework is more CPU time consuming than a standard classical fit but turned out to be more robust and required less human intervention.
In addition it is insensitive to the initialization point as there is no such concept in  nested sampling and provides the full posterior distribution for all parameters, which is out of reach with traditional fitting techniques. In practice, the sampling for a given set of spectra (joint fit of 9 spectra for a given region) takes about 3 hours on 32 cores for a model with 19 degrees of freedom. 

\begin{figure*}
\centering
\includegraphics[scale=0.95, trim = 22 375 0 55, clip=true]{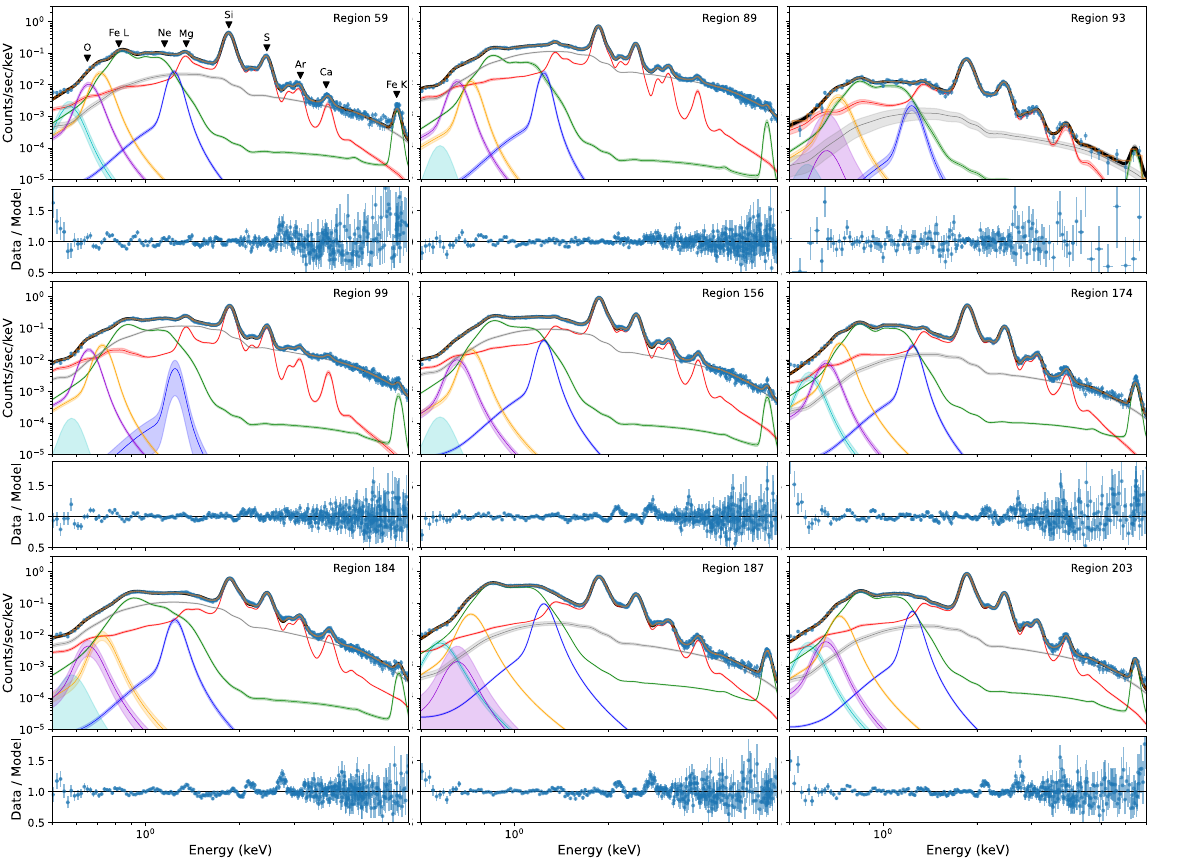}
\caption{\footnotesize Examples for nine regions of spectra and best fits obtained with the Bayesian analysis framework BXA. Each component has its associated uncertainty band at 90\% confidence level (hardly visible for the dominant components). The VNEI for the IMEs is in red, the hotter VNEI for the iron-rich ejecta in green, and the power law for the synchrotron emission in grey. The four Gaussian components are shown in blue (G1 at 1.24 keV), in yellow (G2 at 0.72 keV), in violet (G3 at 0.654 keV), and in cyan (G4 at 0.57 keV).
The expected emission lines of elements are indicated on the first spectrum.
The spectra of the nine Chandra observations are merged for visualization purposes. The associated posterior corner plots for some of these regions are shown in Appendix \ref{Appendix:CornerPlot}. }
\label{fig:ExSpectre}
\end{figure*}

\section{Results}
\label{Section:Results}

Based on a detailed Bayesian spectral study of 211 regions and homogeneous line-of-sight velocities, we obtain detailed maps of 19 physical parameters of the thermal ejecta and non-thermal emission in Tycho's SNR. We provide the thermodynamic state of both the shocked intermediate mass elements and shocked iron-group ejecta. Some of these physical parameters have never been mapped before for Tycho. For each map, an indicative uncertainty is given. It is the median of the uncertainties over all regions for the parameter. Maps of relative error for each parameters are given in Appendix \ref{Appendix:ErreurRelative}. Note that these uncertainties are statistical, the total uncertainties are dominated by the systematical ones.

After discussing the quality of the fits, this section presents the complete ejecta mapping of Tycho's SNR ejecta. We will describe in details these maps and their implications in the next subsections, each dedicated to a given quantity or component.

\subsection{Examples of spectra and their fitting}
\label{Subsection:ExRegions}

To illustrate the variety of spectra and physical properties in Tycho's SNR, we show in Fig. \ref{fig:ExSpectre} the spectra of nine regions, with their model components in different colours, highlighting  distinctive characteristics depending on their locations (indicated in Fig. \ref{fig:Segmentation}). For visual clarity, the spectra of all 2009 observations have been merged. Uncertainties for each model component are provided. They are derived from the posterior distribution obtained from the Bayesian analysis framework. The corresponding corner plots of some posterior distributions are provided in Appendix \ref{Appendix:CornerPlot}. 

Region 59, located in the southeast, is an example of an iron-rich region, in which the iron (VNEI2) temperature (hereafter k$T_2$) is high (11.6 keV).
Region 89 is a region in the southwest with stripes of non-thermal emission \citep{Matsuda2020, Okuno2020}.
Region 93, near the center, is the segmentation region with the fewer number of photons (4$\times$10$^4$ photons, partly due to its position on the detector gap), the thermal components are nevertheless well constrained. 
Region 99 is located in the eastern rim, near the Balmer filaments observed in the optical band, and includes a synchrotron filament. There, the velocity of the forward shock is slower \citep{Williams2016}, probably due to the interaction with a dense clump.
Region 156 is located in the western border, where the density of the ambient medium is low \citep{Williams2013} and the forward shock velocity fast \citep{Williams2016}. It includes another synchrotron filament.
Region 174, in the northeast, is an example of a zone with oxygen line emission.
In the northwestern rim, we choose region 184 where the temperature of the intermediate mass elements is high (k$T_1 \simeq$ 4.5 keV) and the photon index of the power law steep (2.95).
Region 187 is located in the bright emission arc of iron and IMEs in the north.
Finally, region 203 is located in the north, with a lower iron temperature (k$T_2 \simeq$ 7.6 keV).
This selection highlights some particular zones in Tycho's SNR, that we will discuss in more detail in Sect \ref{Section:Discussion}.

\begin{figure*}[h]
\centering
\includegraphics[scale=0.5, trim = 0 0 0 0, clip=true]{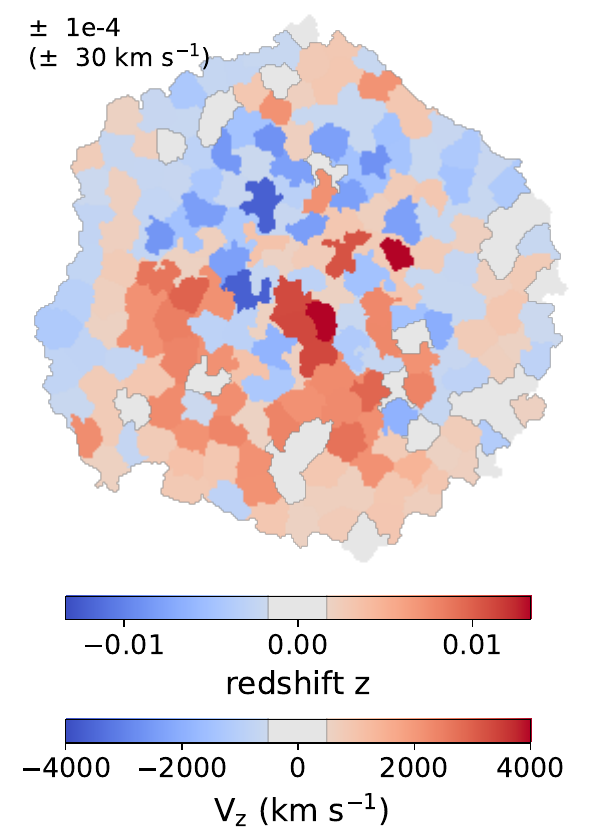}
\includegraphics[scale=0.4, trim = 0 0 0 0, clip=true]{ 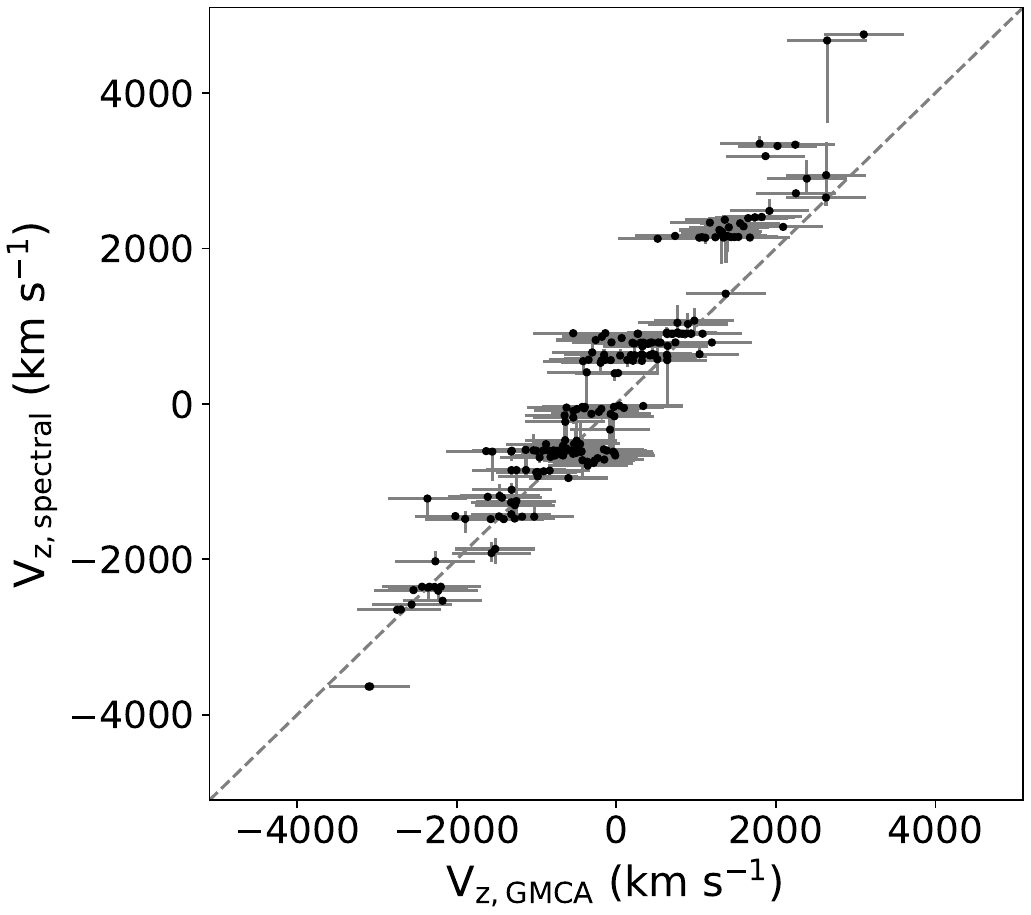}
\includegraphics[scale=0.5, trim = 0 0 0 0, clip=true]{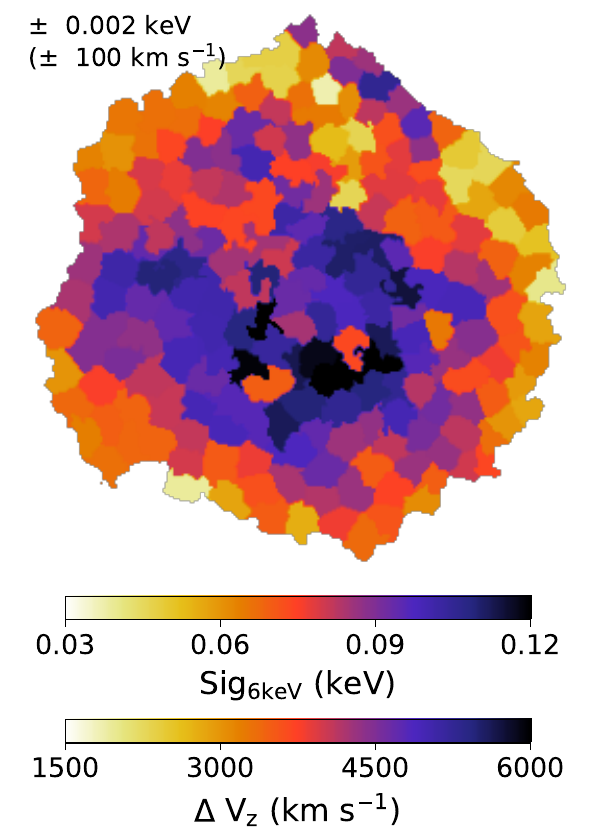}

\caption{\footnotesize \emph{Left:} shocked ejecta velocity $V_{\rm z}$ from our Bayesian spectral study. The uncertainty (top left) only represents statistical error. \emph{Middle:} comparison for each region between the median of the velocity $V_{\rm z}$ in each region from the GMCA maps and the values obtained in this study with a Bayesian spectral fit. \emph{Right:} Sig$_{\rm 6keV}$ parameter, which corresponds to the spectral Gaussian smoothing applied to all the ejecta components (first color bar) equivalent to shocked ejecta velocity dispersion $\Delta V_{\rm z}$ (second color bar).}
\label{fig:Dynamics}
\end{figure*}

As shown in Fig \ref{fig:ExSpectre}, our model reproduces well a range of spectral properties with good fitting results for most of the spectra. 
The corner plots of the posterior distributions for some of these nine regions are provided in Appendix \ref{Appendix:CornerPlot} and show that the parameters are very well constrained, without multimodal distribution. We verified that the priors are in general wide enough that they do not influence the posteriors. We have thus no notion of initialization.
The maps of relative errors (see Appendix \ref{Appendix:ErreurRelative}) have low values and their spatial distributions are decorrelated from the physical structures observed in the parameter maps in the following sections. 

However, we note that in several regions, some spectral residual features are observed. This is the case in the Si-S inter-lines energy band for regions in the northeast. 
In their appendix, \cite{Uchida2024} raise the issue of the sacrificial charge effect in Chandra data. During the readout of a chip (a column of a CCD camera), the transfer of photons suffers a loss. This Charge Transfer Inefficiency (CTI) is known and corrected in the standard reduction process of Chandra. However when several photons fall in the same column during the same temporal frame, the first photon leaves some charge in the trap during readout and the transfer of the second photon will be more efficient because of to the residual charge. As the second photon suffered less loss than the first one, the CTI will over-correct the charge. This process is named the sacrificial charge effect.
We investigate this problem in the deep observation 10095 used in this study. In total, 6\% of the photons are in this case, in a column with several events during the same frame. The effect is more important for the high-count-rate regions: around 11\% of the photons are concerned in the zone corresponding to our regions 187, 173, 179, 161, 165, 149, 139, 131, 156, 146.
By comparing spectra with only the "first" photons arriving in the column during each frame and the total events, we observe that the profiles of the silicon and sulfur lines are slightly modified. They become asymmetrical, wider on the blueshifted side, as we can see in Fig \ref{fig:ExSpectre}. Thus it might partially explain the Si-S interline spectral residuals that we observe, which might affect the measured parameters in the listed regions.
Nevertheless, this effect seems to affect only 10 regions out of the 211 studied here and does not affect the general conclusions of this study.

\subsection{Line-of-sight velocity and dispersion of the shocked ejecta}
\label{Subsection:Vz}

The line-of-sight velocity is obtained from the redshift parameter of our model, which is assumed to be the same for the two thermal components and the four gaussians. Note that the silicon line has the largest impact on the redshift determination.

Figure \ref{fig:Dynamics} presents the $V_{\rm z}$ map (left panel) obtained from our Bayesian spectral study. The large-scale north-south velocity asymmetry is recovered. The local $V_{\rm z}$ structures, not only the north-south dipole, seem to be rather robust as a very similar spatial distribution has been observed in five different ways : on bright knots with Chandra CCDs \citep{Williams2017, Sato2017a} and gratings \citep{Millard2022}, with the GMCA reconstruction on Chandra data \citep{Godinaud2023}, a parameter mapping with XMM-Newton data \citep{Uchida2024}, and this study.
However we note that the XMM-Newton $V_{\rm z}$ map has a smaller spread with maximum values of 2000 km s$^{-1}$. 
To further quantify the correlation between the $V_{\rm z}$ map in this study and that in Fig. \ref{fig:Segmentation}, we compare the median value for each region of the $V_{\rm z}$ obtained with the GMCA method with the $V_{\rm z}$ obtained in this spectral study through the redshift parameter. The correlation is very good, as seen in Fig.~\ref{fig:Dynamics} (middle panel).

The line-of-sight velocity dispersion of the shocked ejecta is derived from the parameter Sig$_{\rm 6keV}$, which describes the spectral Gaussian smoothing that is applied to the thermal components and four Gaussian lines. 
Figure \ref{fig:Dynamics} (right panel) provides this parameter map, expressed in keV, with the correspondance to the line-of-sight equivalent velocity dispersion in km s$^{-1}$.
The line-of-sight ejecta velocity dispersion  appears larger in the center than along the edge.
It is expected as this parameter represents the broadening of the lines, which is mainly due to the opposite velocities between the front and rear shells in the line of sight. To a smaller extent, the line-of-sight dispersion of the physical quantities in each shell causes also a widening, as well as the effect of the thermal broadening.
The highest values, found in the center of the SNR, correspond to a velocity dispersion of $\simeq 6000$ km s$^{-1}$. This is consistent with the maximal values of the line-of-sight velocity $V_{\rm z}$ obtained in this study and in \cite{Godinaud2023}.

\subsection{Absorption parameter $N_{\rm H}$}
\label{subsection:Nh}

\begin{figure}
\centering
\includegraphics[scale=0.62, trim = 0 0 0 0, clip=true]{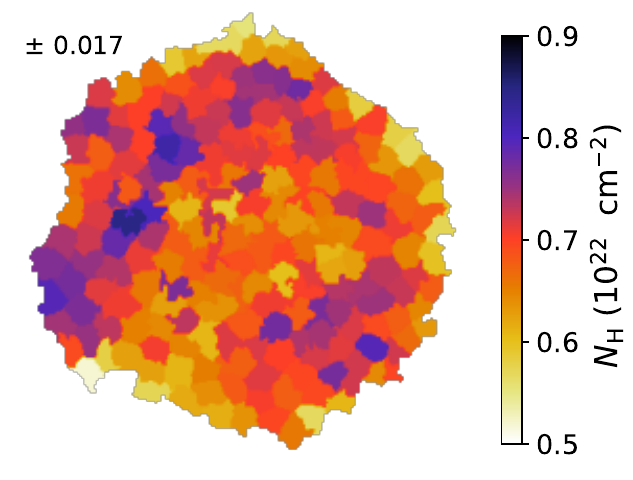}
\caption{\footnotesize Map of the X-ray column density $N_{\rm H}$. The value in the upper-left indicates the mean uncertainty (the median of the uncertainties over all regions).}
\label{fig:CartoNh}
\end{figure}

Fig \ref{fig:CartoNh} presents the map of X-ray column density $N_{\rm H}$ towards Tycho's SNR, derived from our spatially-resolved spectral analysis.
The absorption was assumed to be the same for all our model components (IME, Fe, Gaussian lines, power law, see Table \ref{table:Model}). We find a median value of the absorption parameter $N_{\rm H}$ of $\sim 7\times 10^{21}$ cm$^{-2}$ with a standard deviation between the regions of $\sim 0.6 \times 10^{21}$\,cm$^{-2}$ and a median uncertainty of $0.17\times 10^{21}$\,cm$^{-2}$. These values are in good agreement with previous estimates for Tycho's SNR \citep{Hwang2002, Sato2017a}.

Larger absorption is globally observed on the eastern shell than in the center and western shell, although there are a few locally enhanced absorption regions in the southwest.
This map does not show any correlation with the $V_{\rm z}$ map asymmetry . 

\subsection{Thermal components}
\label{subsection:VNEIs}

\begin{figure*}[h!]
\begin{subfigure}{1\textwidth}
\centering
\includegraphics[scale=0.4, trim = 0 0 0 0, clip=true]{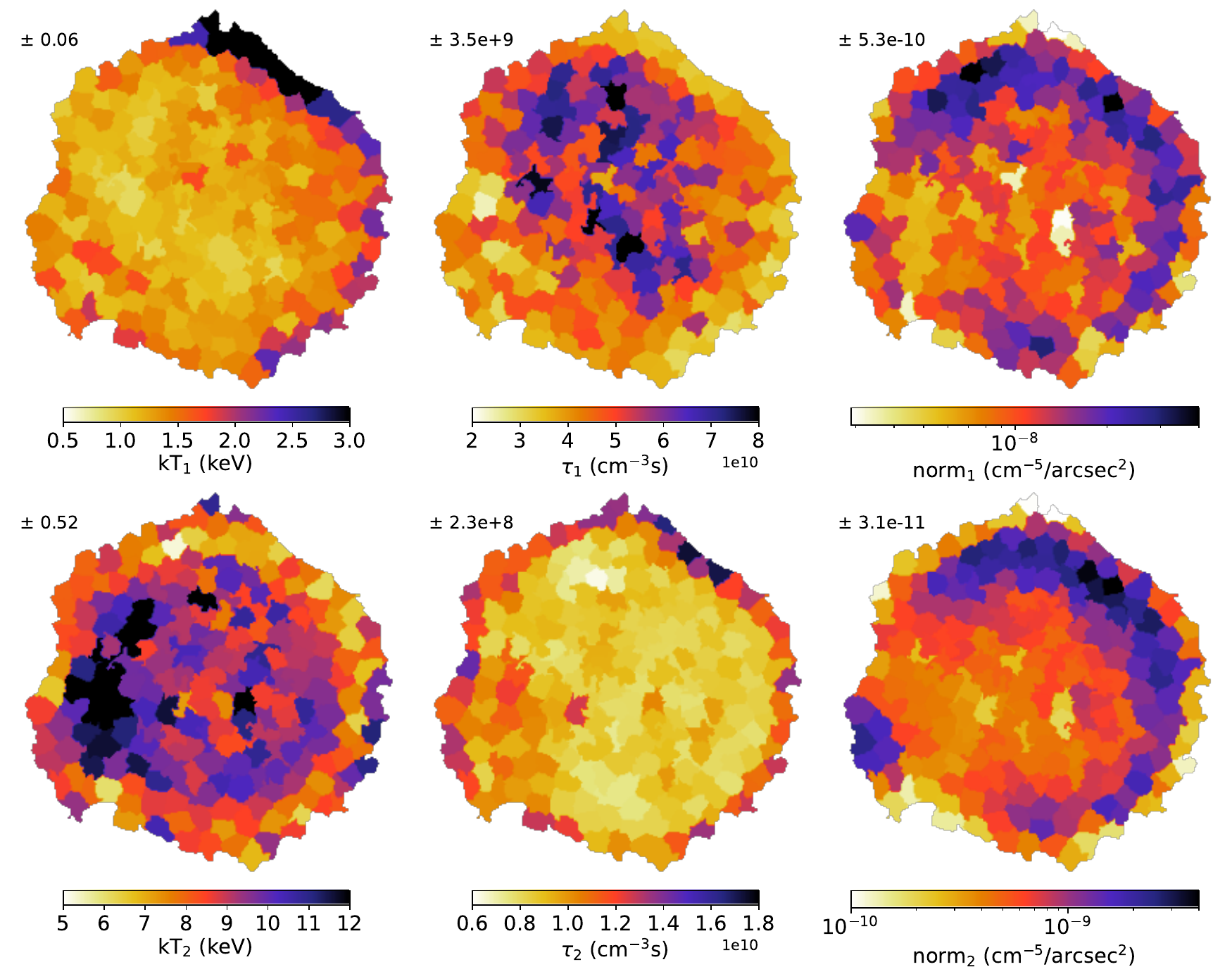} 
\caption{\footnotesize Maps of the main parameters of the two thermal components: temperature, ionization time, and normalization. The normalization is the volume emission measure as defined by VNEI model divided by the region area. The top row corresponds to the VNEI 1 associated with the IMEs, the bottom one is the hotter VNEI 2 related to the iron-rich ejecta. The scales are not the same in both rows.}
\label{fig:CartoVNEI}
\end{subfigure}

\begin{subfigure}{1\textwidth}
\centering
\includegraphics[scale=0.4, trim = 0 0 0 0, clip=true]{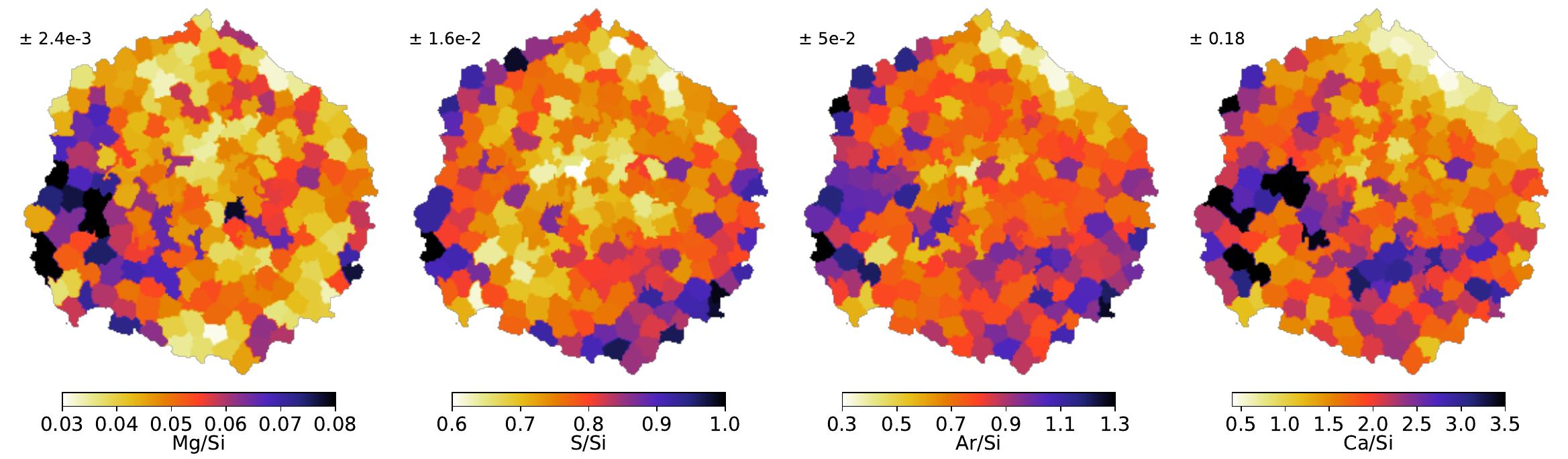}
\caption{\footnotesize Maps of the abundances Mg/Si, S/Si, Ar/Si, and Ca/Si (Si was fixed to 1000). The scales are not the same.}
\label{fig:CartoAbondances}
\end{subfigure}

\begin{subfigure}{1\textwidth}
\centering
\includegraphics[scale=0.4, trim = 0 0 0 0, clip=true]{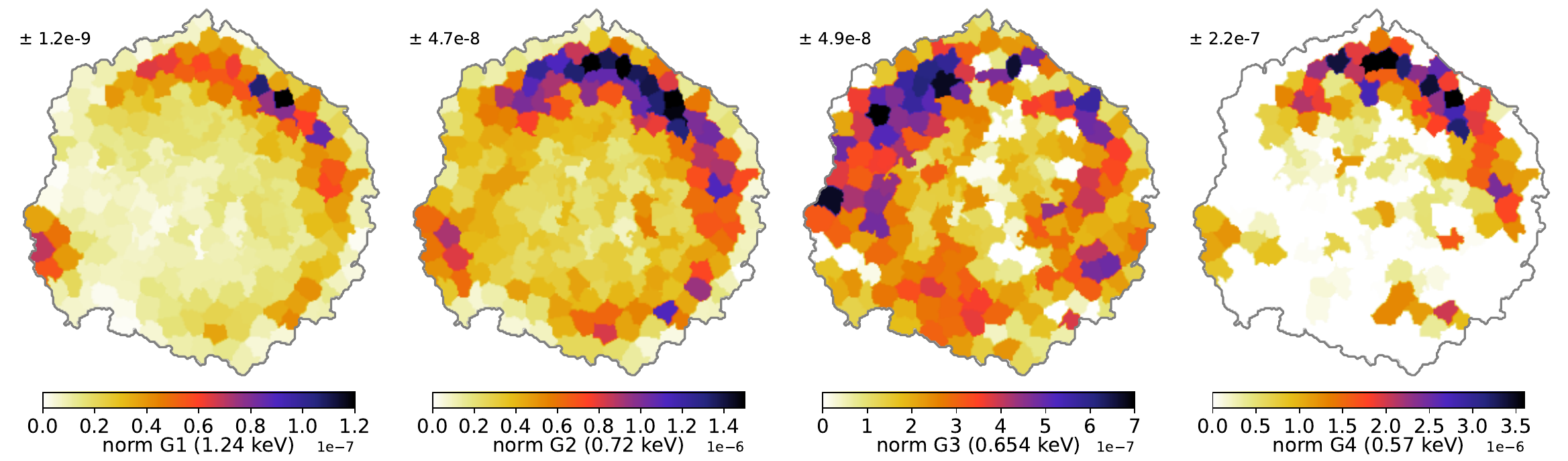}
\caption{\footnotesize Maps of the normalization, divided by the surface of each region, of the four Gaussians added to the model: G1 at 1.24 keV,  G2 at 0.72 keV, G3 at 0.654 keV and G4 at 0.57 keV. The scales are not the same. While G1, G2 and G4 seem correlated with the IME and iron components, the spatial distribution of G3 is very different and is likely associated with O VIII. This would represent the first detection and mapping of oxygen emission in Tycho's SNR. }
\label{fig:CartoNormes}
\end{subfigure}

\caption{\footnotesize Maps of the parameters related to the ejecta emission. The value in the upper-left of each map indicates the mean uncertainty (the median of the uncertainties over all regions)}
\label{fig:Ejecta}
\end{figure*}

Fig \ref{fig:CartoVNEI} presents the electronic temperature k$T$, ionization time $\tau$ and normalization $norm$ of the thermal non-ionization equilibrium components for the IME (top panels) and the Fe-rich (bottom panels) ejecta, derived from the Bayesian spatially-resolved spectral analysis.

The normalization map of the IME shocked ejecta is very similar to the X-ray flux map in the energy band 1.6-4.2 keV, where the emission from Si, S , Ar and Ca lines dominates (see Fig. \ref{fig:App1}, left panel, in Appendix \ref{Appendix : Segmentation}). A bright emission arc extending from the northeast to the southwest of the remnant is visible, with the brightest emission peaking in the northwestern rim. Patchy emission is observed as well in the inner region and in the south-eastern ejecta fast knots. A silicon-rich knot is observed in this region (region C1 in Fig. 2\&3 of \cite{Decourchelle2001} with XMM-Newton observations, or zone B in Fig. 8 of \cite{Yamaguchi2017} with Chandra). This south-east Si-rich knot corresponds to our region 86 and is visible in violet color in the $norm_1$ map of Fig \ref{fig:CartoVNEI}.

The normalization map for the shocked iron-rich ejecta also displays a bright ring of emission from the northeast to the south-west rim, which lies slightly inner to the IME emission \citep{Decourchelle2001} as can be seen from the flux maps (see Fig. \ref{fig:App1} in Appendix \ref{Appendix : Segmentation}). The inner regions appear fainter and rather uniform, compared to the IME normalization map.
The known iron-dominated fast knots in the south-east are clearly visible. 

Focusing on the IME component, the temperature (k$T_1$) map is mostly homogeneous in the inner and eastern regions with values between 0.9 and 1.5 keV. The regions along the rim in the western half are hotter, typically more than 2 keV with some extremes in the northwest. Conversely, the ionization time $\tau_1$ has a smaller value along this western edge, of about 3 $\times10^{10}$ cm$^{-3}$ s and becomes patchy in the inner regions with values up to 8 $\times10^{10}$ cm$^{-3}$ s.

For the iron-rich ejecta component, we introduced a Gaussian prior on the temperature parameter to overcome the degeneracy between temperature k$T_2$ and ionization time $\tau_2$, observed along the rim in the corner plots of the posterior distributions due to faint (or absence of) Fe-K line emission. 
The final map of k$T_2$ is complex: along the rim, the temperature is around 6-8 keV and becomes higher in the center (8-12 keV). In particular, there is an eastern arc with values up to 12 keV. The ionization time map $\tau_2$ appears rather homogeneous with low values in the inner regions, around 0.7 - 1 $\times10^{10}$ cm$^{-3}$ s. The ionization time is higher along the rim, from 1.2 up to 1.8 $\times10^{10}$ cm$^{-3}$ s.

\begin{figure}
\centering
\includegraphics[scale=0.5, trim = 0 0 0 0, clip=true]{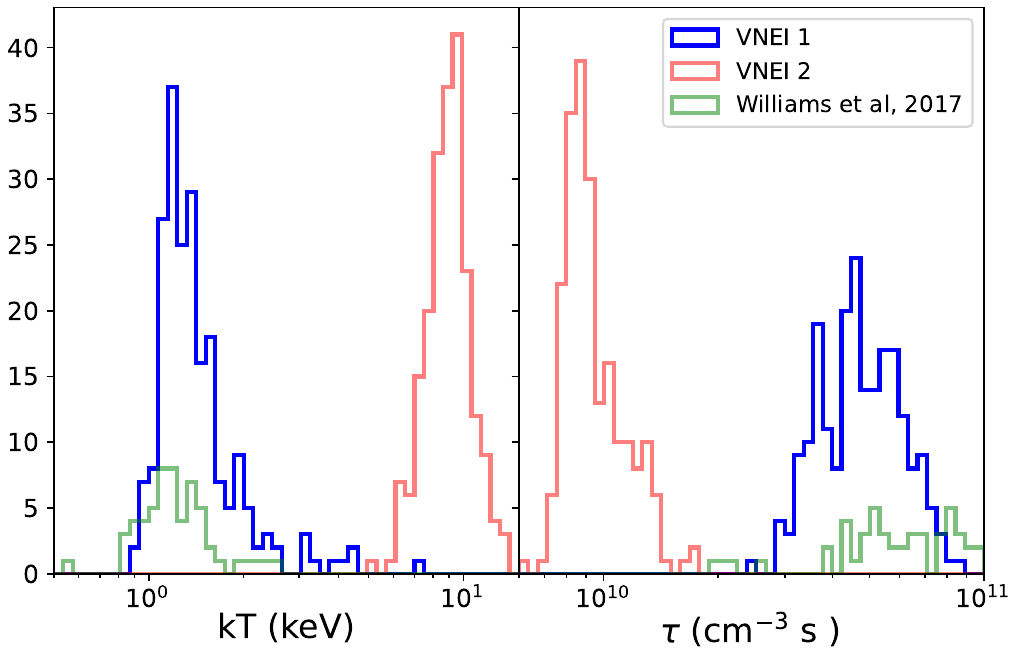}
\caption{\footnotesize Histograms of the temperature and ionization time for the shocked IME (blue, VNEI 1) and iron-rich (red, VNEI 2) ejecta in Tycho's SNR from this study. In green, for comparison the histogram for 57 bright individual ejecta knots fitted in the IME energy band in \cite{Williams2017}.}
\label{fig:hist_kT_Tau}
\end{figure}

Fig. \ref{fig:hist_kT_Tau} presents the histograms of the temperature and ionization time for both components of the shocked ejecta (IME and iron) in Tycho's SNR.
The spread in temperatures and ionization times for the 57 bright knots fitted by \cite{Williams2017} in the IME energy band, is similar to that obtained for the IME ejecta component from our global study. It was not obvious that bright individual knots would have similar properties to the global IME ejecta.

\subsection{Spatial distribution of the abundances}
\label{subsection:Abundances}

Fig. \ref{fig:CartoAbondances} presents the IME abundance maps of Mg, S, Ar and Ca, relative to Si.
Their spatial distributions are relatively uniform, with some enhanced abundances in specific locations.
The Mg/Si map shows weak abundance of Mg relative to Si (below 10\%) and variations by less than a factor two over the remnant. The higher values are located in the southeastern sector and around the known iron-rich knot. The maps of S/Si and Ar/Si indicate S and Ar abundances similar to Si, with variations of less than a factor two for S and up to a factor of about 4 for Ar. The highest values are observed in the southwestern, eastern and northeastern edges of the SNR. For Ca/Si, the map indicates values of Ca ranging from half to a few that of Si. The highest values are observed in the south-eastern region, with a possible structure extending through the center to the south-western rim.

We remark that the S, Ar and Ca maps seem to be correlated, and the Mg map much less. The first ones are clearly associated to the ejecta, but the question is open for Mg, which can also arise from circumstellar material, coming from the progenitor system. This will be discussed in Sect. \ref{subsection:Abondances} in relation with nucleosynthesis model expectations. 

The low abundances found at the north-western edge in the S/Si, Ar/Si and especially the Ca/Si maps must be interpreted with caution.
In these regions, the IME temperature is found to be very high and it may impact the abundance determination of the heaviest IMEs as we use only one temperature for the four elements.

\subsection{On the origin of the additional Gaussians}
\label{subsection:Gaussians}

Fig. \ref{fig:CartoNormes} presents the normalization maps of the four Gaussian lines added to account for features in the residuals, which were not reproduced by the two VNEI components: G1 at 1.24 keV,  G2 at 0.72 keV, G3 at 0.654 keV, and G4 at 0.57 keV. The Gaussian components G1 and G2 are well constrained and contribute in almost all regions. This is also the case for the G3 component, except in certain predominantly central regions, while the G4 component is only present in the brightest regions. 
Their respective contributions are illustrated in Fig. \ref{fig:ExSpectre}, with examples of spectra from different regions and properties. 
Maps of G1, G2 and G4 have a similar morphology, which is very close to the distribution of iron-rich ejecta (VNEI2 norm, see Fig. \ref{fig:CartoVNEI}), while component G3 shows a completely different morphology.

\begin{figure}
\centering
\includegraphics[scale=0.4, trim = 0 0 0 0, clip=true]{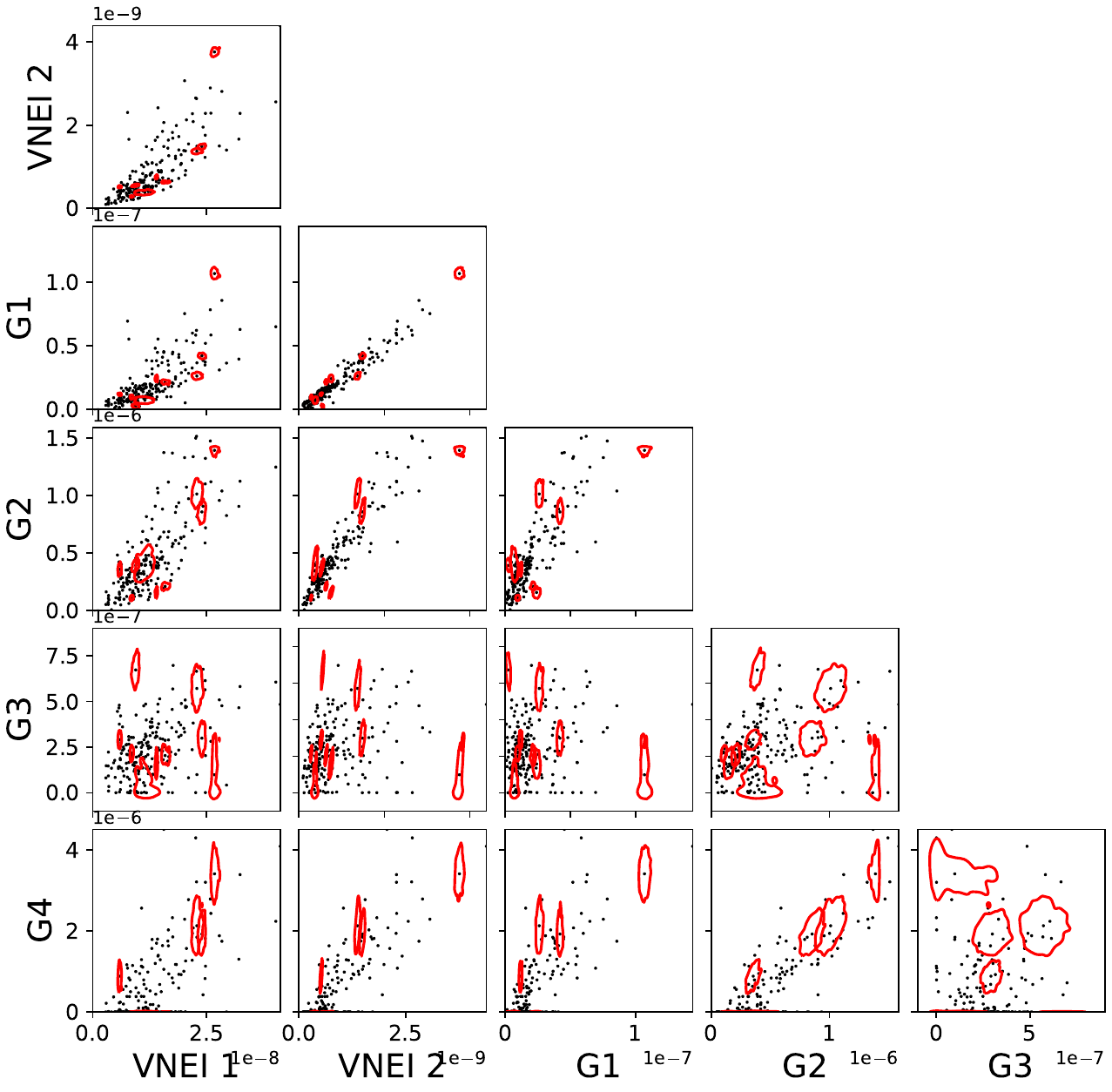}
\caption{\footnotesize Correlation between the normalizations, divided by the surface of the region, of the two thermal components and the four Gaussian components. We add in red the contours of uncertainty at 1 sigma for the nine example regions of Fig \ref{fig:ExSpectre}.}
\label{fig:CorrelationNormes}
\end{figure}

To quantify the level of correlation between the Gaussian line maps and the iron-rich and IME components, we present in Fig. \ref{fig:CorrelationNormes} the scatter plots between all component normalizations related to thermal emission. We illustrate the level of constraints on these normalizations, by including uncertainty contours for the nine regions whose spectra are shown in Fig. \ref{fig:ExSpectre}.

The good correlations between the normalizations of the Gaussian components G1 and G2 (and G4 to some extent) and that of the iron-rich VNEI confirm that these lines might be associated with iron emission. 
Indeed, iron lines are expected at 1.24 keV, such as Fe XXV for a 8 keV plasma temperature (or Fe XXI for a 1 keV plasma temperature), and at 0.72 keV, such as Fe XXIV for a plasma temperature of 8 keV (or Fe XVII for a 1 keV temperature plasma). The component G4 was initially expected to represent the O VII line at 0.57 keV (for a plasma of 1 keV) but its cartography and normalization scatter plot seem instead well associated with that of the Gaussian G2 and Fe. In AtomDB, there are possible iron lines at this energy (e.g, a Fe XXIV line for a temperature of 8 keV).
The intensities of those lines might be underestimated in the AtomDB database, or they could come from colder iron ejecta that we did not model, explaining why these lines do not appear in our the iron-rich VNEI.

The Gaussian G3 component at 0.654 keV shows a morphology clearly different from that of the three other Gaussian components and from that of IME and iron-rich ejecta. Its emission peaks in the northeastern rim and extends towards the south and the west, with little emission in the northwestern rim, where all other components are the brightest. There is also no correlation between its normalization and that of the others in Fig. \ref{fig:CorrelationNormes}. At this energy, oxygen is clearly the dominant emission, whatever the plasma temperature (see AtomDB). We therefore attribute this G3 component to oxygen emission, and interpret this as the first detection and mapping of oxygen in Tycho's SNR. The question of its origin (ejecta, CSM or ISM) will be discussed in more detail in Sect. \ref{subsection:Abondances}.

\subsection{Synchrotron emission}
\label{subsection:Syn}

\begin{figure*}
\centering
\includegraphics[scale=0.8, trim = 0 0 0 0, clip=true]{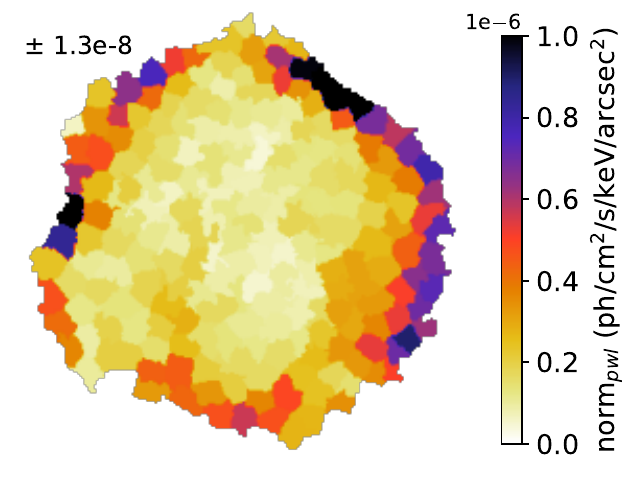}
\includegraphics[scale=0.4, trim = 0 0 0 0, clip=true]{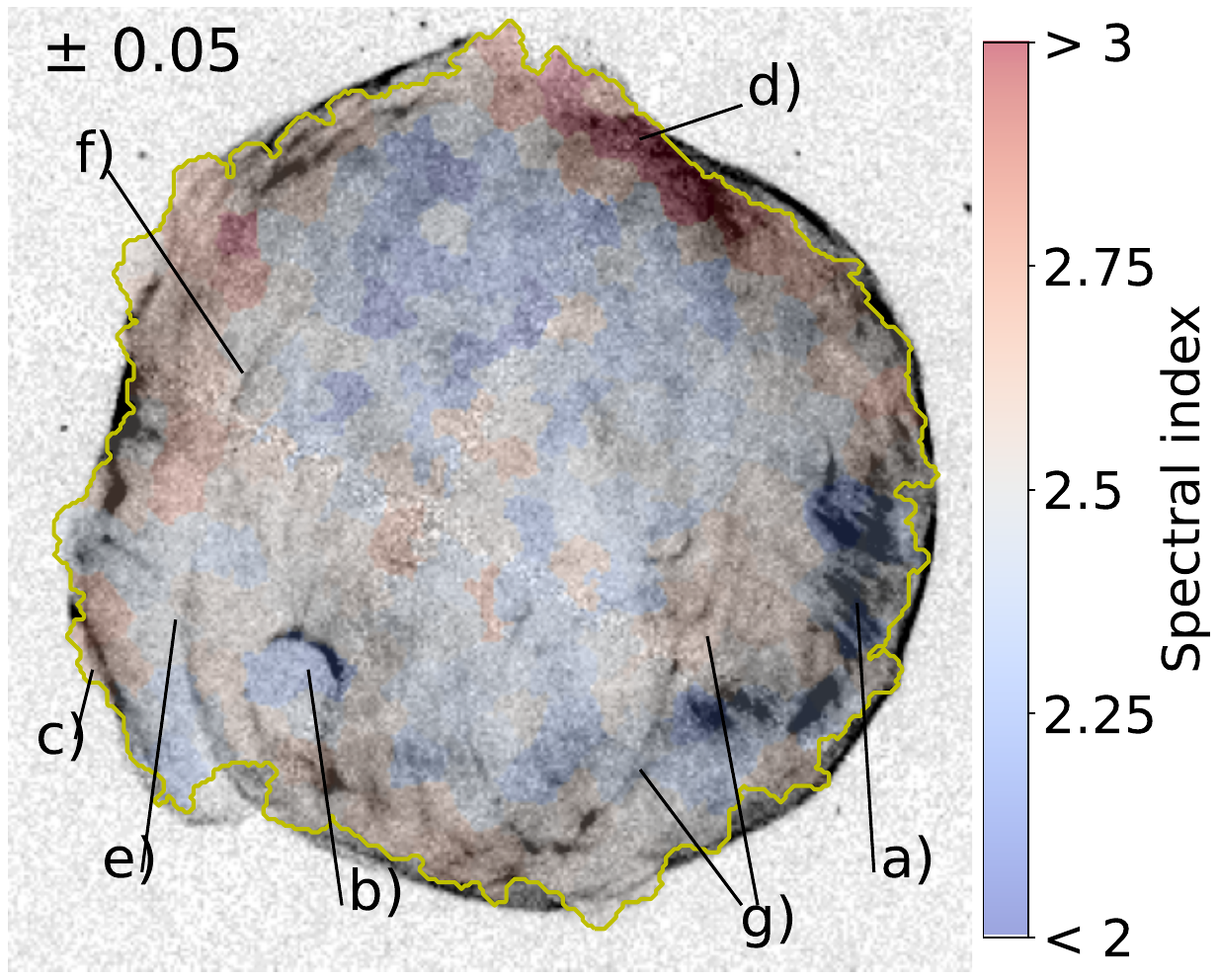}
\caption{\footnotesize Left: Map of the norm of the power law component. Right: Map of the photon index of the power law component in color overlaid on the map of the emission between 4 and 6 keV, associated with the synchrotron emission. The background image is corrected from the exposure map. The yellow line traces the outer contour of the ejecta and our regions (see Fig \ref{fig:Segmentation}). Some features are highlighted and described in the text (Sect. \ref{subsection:Syn}).  } 
\label{fig:Sync}
\end{figure*}

Figure \ref{fig:Sync} presents the maps of parameters related to the non-thermal component (modeled as a power law), which is interpreted as synchrotron emission. Both its photon index and normalization are well constrained in all regions. In most previous studies, the analyses of the non-thermal emission were focused on the rims. For the first time, we present a complete mapping of the diffuse synchrotron emission within the SNR.

The power law normalization map, seen in Fig. \ref{fig:Sync} (left panel), is well correlated with the X-ray emission between 4 and 6 keV (right panel) where the non-thermal emission dominates, with the brightest regions at the rim. It also has a morphology reminiscent of the radio emission \citep[see Fig. 1 of][]{Dickel1991}.

Fig. \ref{fig:Sync} (right panel) shows the photon index map of the power law (in color), superimposed on the 4-6 keV emission (black-and-white background). The photon index varies from values of about 2 to above 3 in the north-western rim, with a median value of 2.5 which is consistent with the bibliography \citep{Cassam-Chenai2007, Tran2015}. It shows various regimes in different regions of the SNR that we bring to the fore with letters in Fig. \ref{fig:Sync}.

The non-thermal stripes (denoted \emph{a}) in the southwest have a low photon index, between 2.1 and 2.6, in agreement with \cite{Matsuda2020}, who found values between 2.06 and 2.55. They interpret these values as an effect of an amplified magnetic field and/or some peculiar regime of particle acceleration.
\cite{Lu2011} fitted the non-thermal arc in the southeastern quarter (denoted \emph{b} ) and found a photon index of 2.47. The equivalent regions in our segmentation (48 and 45) have values around 2.2. In that article, this structure is proposed to be due to the interaction between the ejecta and the envelope of the companion star.
The filament marked \emph{c} at the shock, near the iron-rich knot, was fitted by \cite{Yamaguchi2017}. They found photon indices of 2.58 and 2.68 in agreement with our regions 62 and 43, with respectively 2.77 and 2.67.
The region in the north-western edge (denoted \emph{d}) is again notable with photon index between 2.8 et 3.1. This region will be more discussed in Sect \ref{subsection:Relation kT tau}.
Finally, we observe several faint vertical filamentary non-thermal structures in the SNR, marked as \emph{e}, \emph{f} and \emph{g}. It is hard to study their non-thermal properties, the ejecta emission being dominant in these regions. With this study, we have evaluated for the first time the photon index of several of these sub-dominant filaments in the center of the remnant, at values around 2.5. 

\section{Discussion}
\label{Section:Discussion}

Using the deep ($\sim$ 700 ks) Chandra observation combined with our Bayesian analysis of 211 regions, the parameter maps obtained for Tycho's SNR contain a wealth of information on the emitting conditions in the IME and iron ejecta, as well as on the non-thermal plasma. We highlight in the following sections a selection of scientific results.

\subsection{Large-scale $V_{\rm z}$ asymmetry}
\label{subsection:Assym N/S}

A large-scale $V_{\rm z}$ asymmetry is observed between the north and south in Tycho's SNR, the north being dominated by blueshifted ejecta and the south dominantly redshifted as presented in Fig. \ref{fig:Dynamics}.
The $V_{\rm z}$ map shows an integrated velocity along the line of sight, weighted by luminosity. Therefore, this dipole can be an asymmetry of velocity or/and luminosity. As we do not see flux differences between the north and south (for example in the map of IME normalization in Fig. \ref{fig:CartoNormes}), the hypothesis of an asymmetry of velocity is favored.
Either it can be innate, due to an anisotropy during the explosion that causes large-scale jets, or it can be acquired during the expansion, due to circumstellar structures. An asymmetric slowdown may be due to inhomogeneities in the ambient medium, which can have a torus structure. In both cases we can probe the progenitor history thanks to SNR observations as studied by numerical simulations \citep{Chiotellis2013, Ferrand2019}.

A large-scale interaction with the ambient medium would leave a fingerprint, even centuries later. However, in the maps of temperature and ionization time of the IME ejecta in Fig. \ref{fig:CartoVNEI}, there is no clear north-south asymmetry. These parameters are not correlated with our map of integrated velocity along the line of sight.
However, an interaction with a symmetrical titled overdensity, like an annulus, could explain an asymmetry in velocity but not for the over physical properties. So, based on the cartography shown here, the origin of this large-scale $V_{\rm z}$ asymmetry seems related to early properties either in the explosion or in the immediate CSM.

\subsection{Anti-correlation k$T$ - $\tau$}
\label{subsection:Relation kT tau}

\begin{figure*}
\centering
\includegraphics[scale=0.25, trim = 0 0 0 0, clip=true]{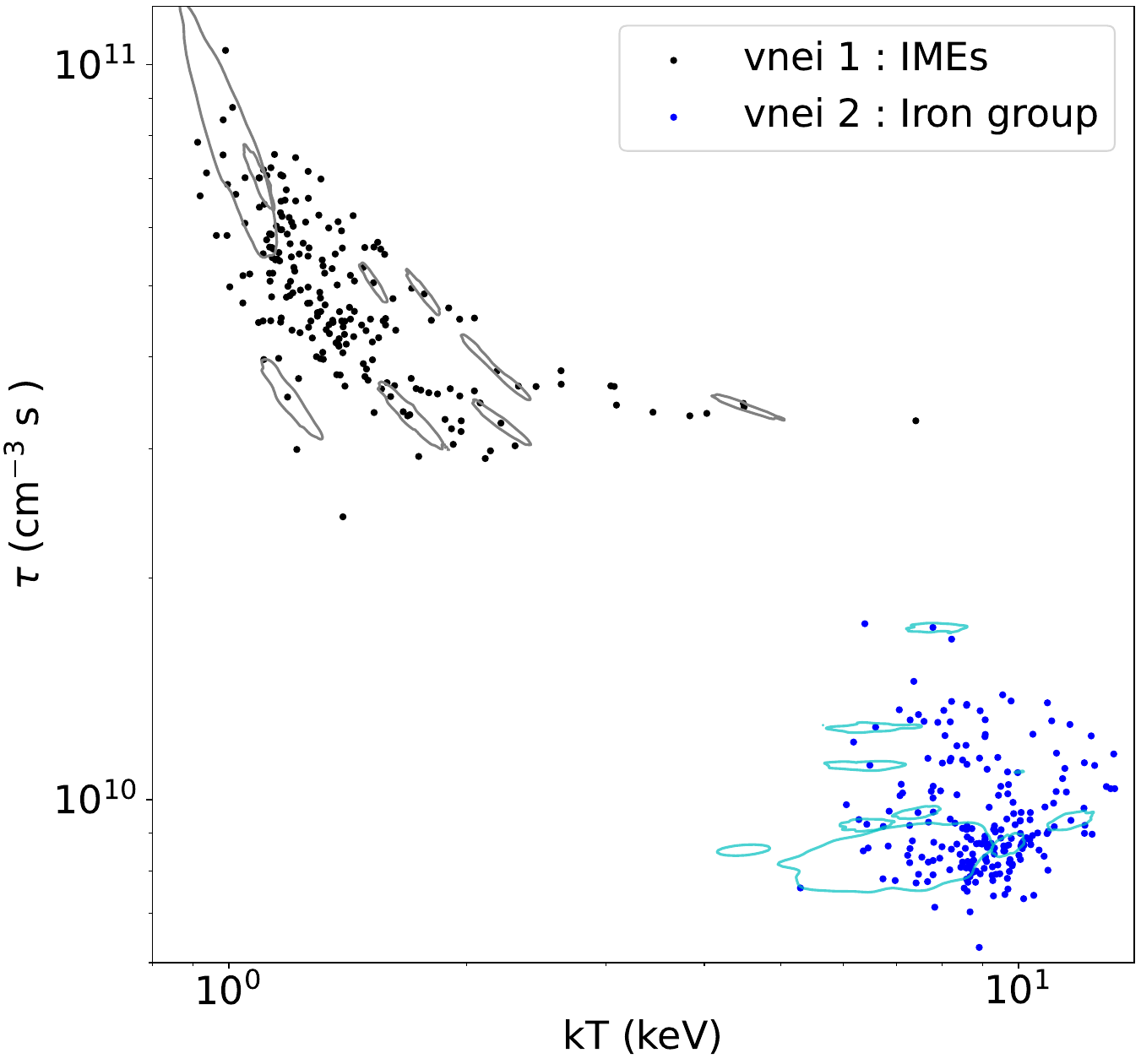}
\includegraphics[scale=0.3, trim = 0 0 0 0, clip=true]{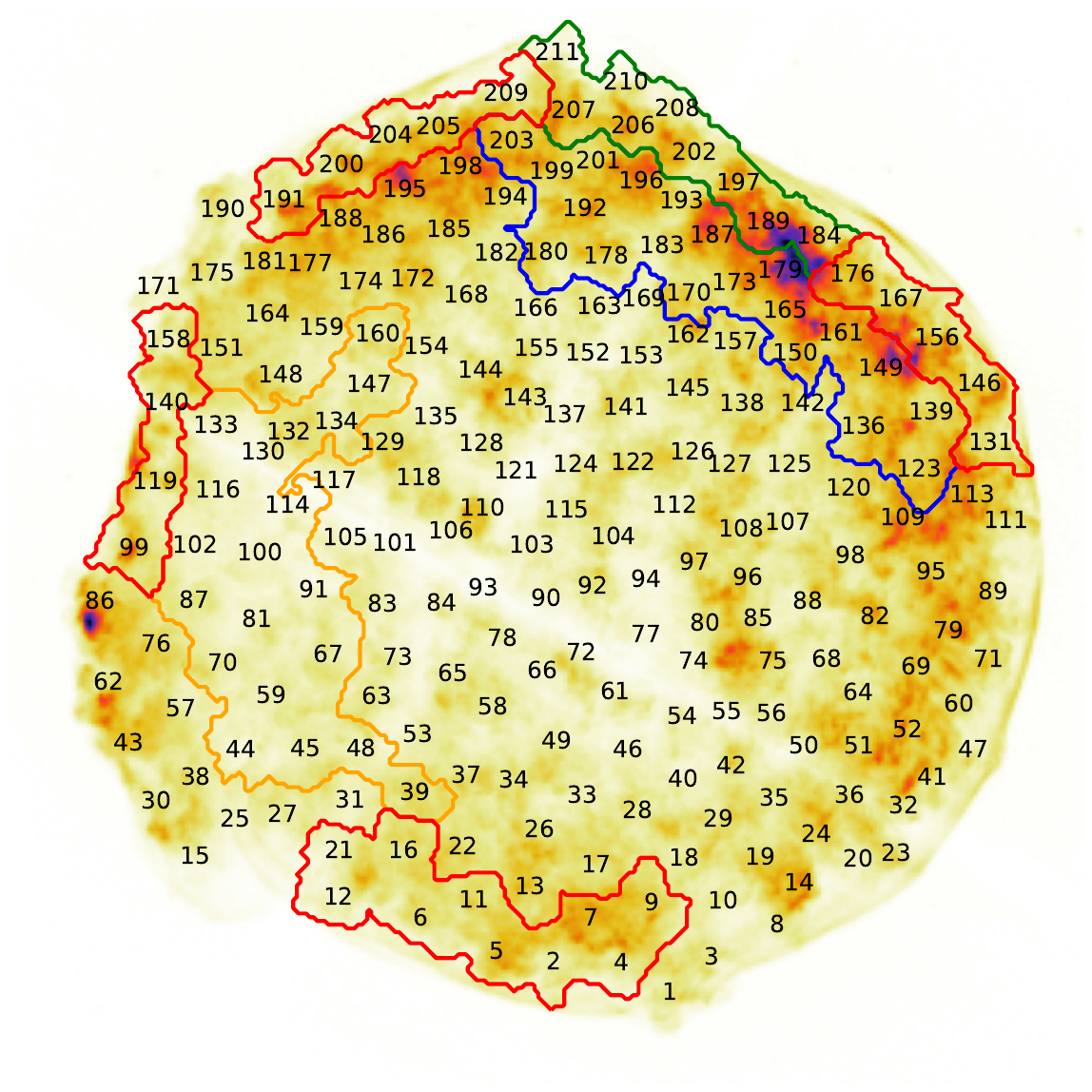} 
\includegraphics[scale=0.25, trim = 0 0 0 0, clip=true]{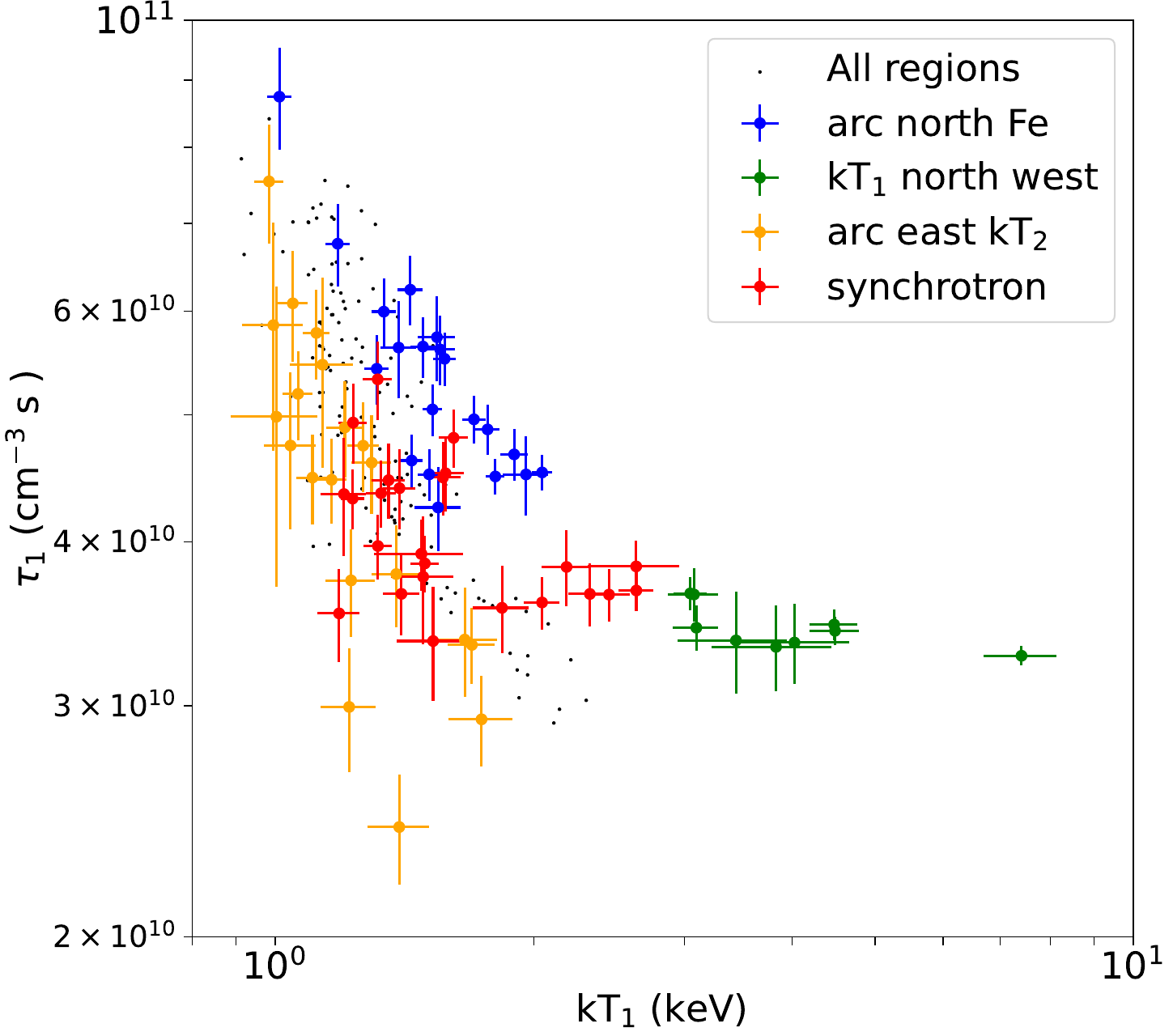}
\caption{\footnotesize \emph{Left panel: }Correlation over the entire SNR between the temperature k$T$ and ionization time $\tau$ for the IME thermal component (in black) and iron-rich thermal component (in blue). We add the contours of uncertainty at 1 sigma for the nine example regions (spectra in Fig \ref{fig:ExSpectre}). \emph{Middle panel:}  Clusters of regions with different physical properties (see text in sect. \ref{subsection:Relation kT tau}) superimposed on the flux map between 0.5 and 7 keV. \emph{Right panel:} Correlation between the temperature and ionization time of the IME thermal component for specific regions (coded with different colors for each cluster). The error bars are the standard deviations of the posterior distributions for each region.} 
\label{fig:relationkT_Tau}
\end{figure*}

An interesting result that we obtain is the relation between the temperature and ionization time, as shown in Fig. \ref{fig:relationkT_Tau}. For the IMEs, we obtain a banana anti-correlation between k$T_1$ and $\tau_1$. For the iron-rich ejecta, no particular correlation is observed. 

This is not the first time that this banana anti-correlation between k$T$ and $\tau$ appears in studies related to supernova remnants. For Tycho's SNR, if we use the values in Table 2 of \cite{Williams2017}, a similar  relation is observed (not shown in the original article). With the \emph{Chandra} data, they fitted 57 bright knots with an absorbed model with a VNEI and a power law in the [1.2, 2.8] keV band, which corresponds to our IME component.
\cite{Sun2019} did a complete cartography of Kepler SNR also with the \emph{Chandra} data, with a VNEI + srcut + missing lines, between 0.3 and 8 keV (with little statistics above 3 keV) and also revealed this correlation.
A similar result was found by \cite{Li2015} for SN1006 with the analysis of XMM-Newton data in the [0.3, 8] keV band (with little statistics above 2 keV). They used a model VNEI + srcut + wabs + a few missing lines. 

So this banana anti-correlation has been observed for multiple SNRs, with different telescopes, and spectral models. The question is to understand whether there is a true physical relation (due to the SNR plasma properties) between the temperature and ionization time or a degeneracy between these parameters.

To constrain k$T$ and $\tau$ by fitting a spectrum, the centroids of the lines, the ratios of lines associated with different levels of ionization, and the continuum are used. There could be some degeneracy with other parameters (like the redshift and the normalization) but also between k$T$ and $\tau$. In the appendix B of \cite{Godinaud2023}, the centroid energy of the Si line (at the Chandra resolution) is shown as a function of these two parameters and we find a similar banana anti-correlation. There is a similar effect in the appendix of \cite{Millard2020} where figure A1 shows some line-emission ratios according to k$T$ and $\tau$. However, there is other information in the spectrum that allows to constrain these parameters. 
In the corner plots obtained with BXA in Appendix \ref{Appendix:CornerPlot}, we have for the first time a general view of the statistical landscape of our fitting model and correlation between all parameters. In each corner plot, we find this anti-correlation between temperature and ionization time. Using the information from the ellipse of uncertainties for nine regions with different properties in Fig. \ref
{fig:relationkT_Tau} (left panel), we can see that most  regions are well constrained and that the spread in the banana shape arises from various conditions in the 211 regions, and is likely not solely due to the degeneracy of each region. 

To investigate the origin of this correlation, and whether it is linked to some particular physical conditions or underlying physics, we add some spatial information to the k$T$-$\tau$ diagram.
In Fig. \ref{fig:relationkT_Tau} (middle panel), we have selected four groups of regions, each presenting specific physical features in the parameter maps. In red, we have grouped the regions where the normalization of the power law component is high. Although our study focuses on the thermal ejecta emission, some synchrotron filaments remain, both at the edge of the remnant and in the south-western interior, where numerous non-thermal stripes are observed (see Sect. \ref{subsection:Syn}).
The bright arc in the north, where the normalization of the two thermal components is maximum, is shown in blue. 
We have marked in orange an arc in the east, where the temperature of the iron-rich component is particularly high (see Fig. \ref{fig:CartoVNEI}). This is the first time that this structure is highlighted. 
Finally, in green, we have identified a feature in the north east where the IME thermal component is the hottest and the value of Sig$_{\rm 6 keV}$ is also abnormally high for an edge (see Fig. \ref{fig:CartoVNEI}).
A correlation appears between these groups of regions and their physical properties in Fig. \ref{fig:relationkT_Tau} right panel). For example, the iron arc in the north (in blue) and the arc in the east  with a higher k$T_2$ (in orange) form two different branches in the k$T$-$\tau$ plot. In mapping Kepler's SNR, \cite{Sun2019} also observed two different branches of k$T$-$\tau$, interpreted as originating from ejecta and CSM emission.
This banana-shape anti-correlation therefore seems to stem from different conditions in the supernova remnant, possibly related to different initial conditions, composition or/and evolution history. However the mechanism leading to this anti-correlation remains unexplained. 

\subsection{Abundances and nucleosynthesis}
\label{subsection:Abondances}

\begin{figure*}
\centering
\includegraphics[scale=0.45, trim = 0 0 0 0, clip=true]{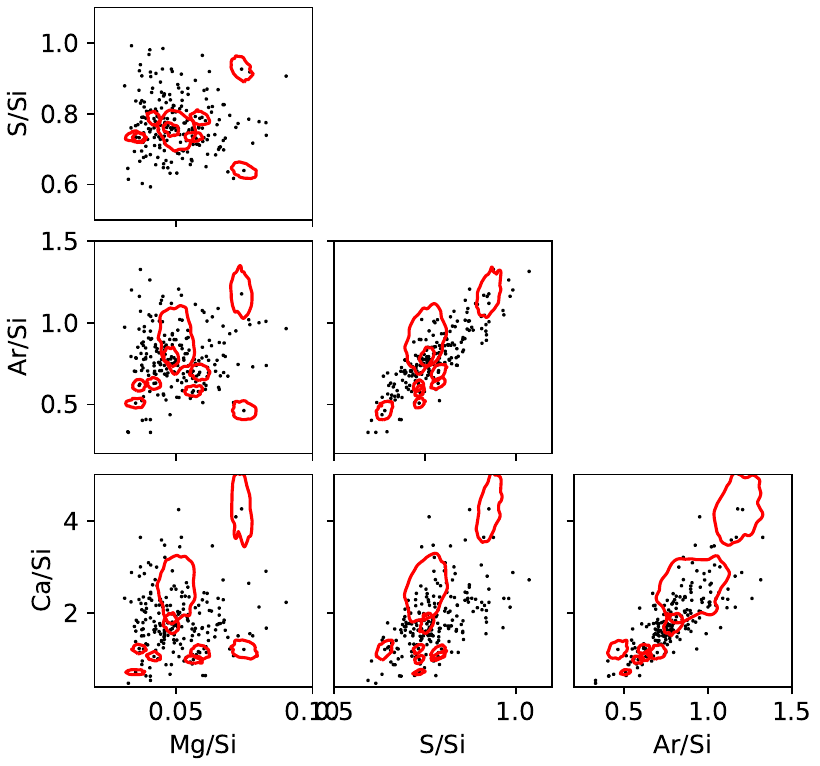}
\includegraphics[scale=0.45, trim = 0 0 0 0, clip=true]{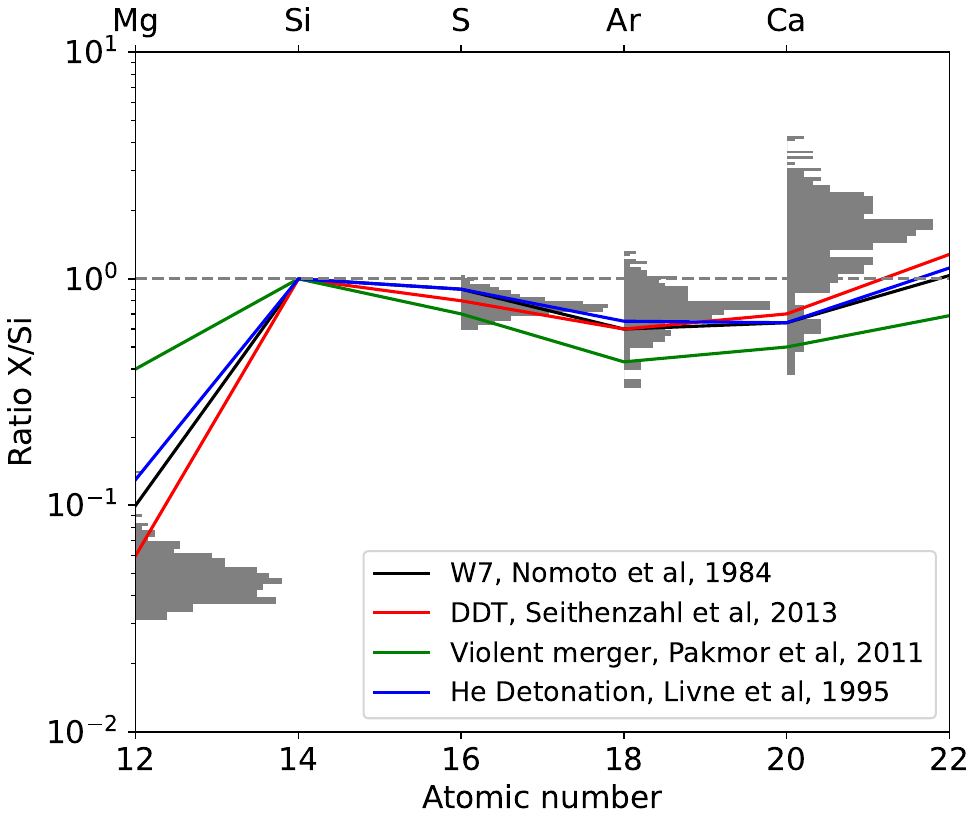}
\caption{\footnotesize \emph{Left panel:} Correlations between the different abundances relative to Si. We add in red the contours of uncertainty at 1 sigma for the nine example regions. \emph{Right panel:} Comparison between our abundance measurements for the 211 regions (the four histograms for each element) and four nucleosynthesis models summarized by \cite{Williams2020}.
}
\label{fig:Nucleosynthese}
\end{figure*}

We provided in Fig. \ref{fig:CartoAbondances} the abundance ratio maps of several elements relative to silicon. Fig. \ref{fig:Nucleosynthese} (left panel) presents the correlation scatter plots between these abundances.
We observe a good abundance correlation between S/Si, Ar/Si and Ca/Si, so they come probably from the same layers during their production. Their abundance maps are indeed similar.
Mg is sometimes associated to the circumstellar medium but it can be also found in the ejecta, it is what we suppose by modeling the Mg line in the IME VNEI. The Mg/Si abundance map is relatively uniform in most of the remnant with values around 0.05, except in the eastern and south-eastern regions, where a few values of up to 0.08 are observed. In Fig. \ref{fig:Nucleosynthese} (left panel), Mg/Si does not correlate with S/Si, Ar/Si and Ca/Si, possibly suggesting a production during the explosion in a different layer.

We can also compare our measurements with nucleosynthesis models of SN Type Ia. Fig \ref{fig:Nucleosynthese} (right panel) summarizes our global abundance measurements and compares them with the bibliography from \cite{Williams2020}. They gathered four nucleosynthesis models of various explosion mechanisms. Our measurements agree in general with their predictions. We exclude the violent merger model that would produce too much Mg, but it would be difficult to differentiate from the three others with only the IMEs. We note that we have more Ca and less Mg than expected.
The resolved mappings of abundances of this study open the door to two or three-dimensional simulations of expected yields to add more constraints on various nucleosynthesis models. For example \cite{Seitenzahl2013} predict a spatial distribution of elements becoming more homogeneous for an increasing number of ignition points in a 3D delayed detonation model. Our mapping shows structured abundance maps with overabundant regions, implying at most a hundred ignition points. The position of the Mg region around the iron-rich knot is surprising : could this be related to a stratification around an iron bullet during the explosion ?

In Sect \ref{Section:Results}, we interpret the normalization of the Gaussian G3 as the emission of the O VIII line, mapped for the first time in the Tycho's SNR. This increased level of oxygen is mostly in the North-East part of the SNR. We do not measure the oxygen abundance here and an additional study to retrieve the physical parameters of this component would be needed but we can speculate about its origin. 
This oxygen can be of ejecta origin heated by the reverse shock. \cite{Seitenzahl2013} predict some production of oxygen in the outer regions during the explosion with an asymmetry level depending on the number of ignition points. 
The second option is the heating of the oxygen from the surrounding environment by the forward shock. In this case, there is an emission of oxygen lines in the X-ray band but it is very faint, especially in comparison with the ejecta emission. However, it is a zone with a lot of synchrotron filaments that trace the forward shock. If there is some oxygen in the surrounding medium, it is expected to emit just after the forward shock, as filamentary thermal structure. Perhaps we measure the accumulation of these thin structures in these regions.

\subsection{Prospective for high-resolution spectroscopy}
\label{subsection:XRISM}

\begin{figure*}
\centering
\includegraphics[scale=0.9, trim = 0 525 0 50, clip=true]{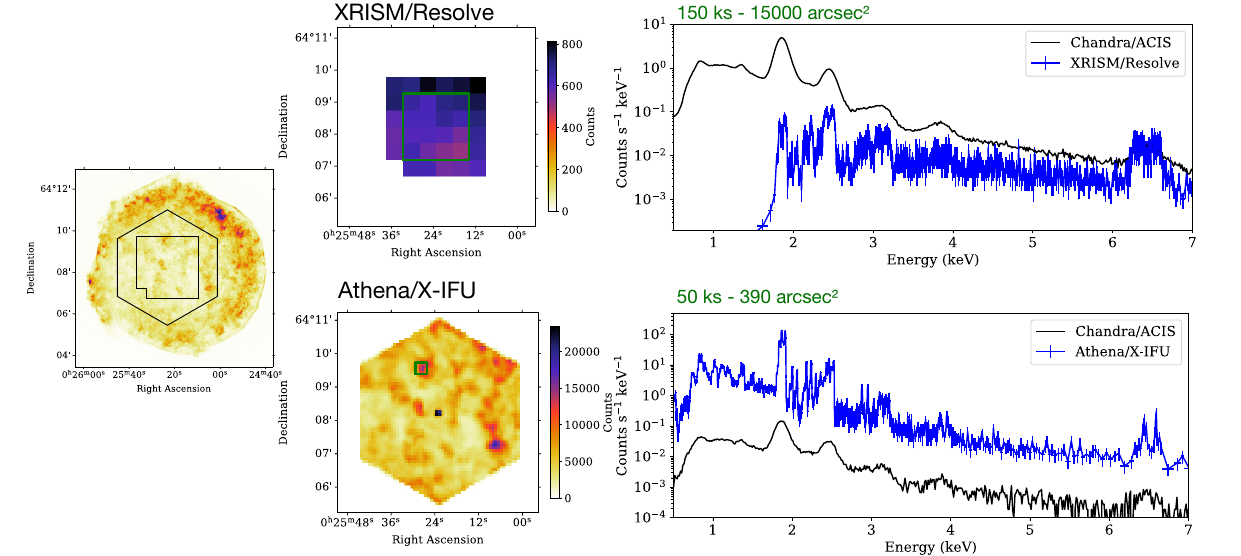}
\caption{\footnotesize SIXTE simulations of the center of Tycho's SNR based on the parameter mapping presented in this study. \emph{Left panel:} The flux image of Tycho's SNR as observed by Chandra, corrected by the exposure map. We overlay the fields of view of the instruments XRISM/Resolve and Athena/X-IFU. \emph{Right upper panel:} Corresponding SIXTE image and spectrum simulation with XRISM/Resolve for 150 ks and the  gate valve closed. The spectrum in blue is extracted from a region of 4$\times$4 Resolve pixels (large green square in the XRISM image). \emph{Right lower panel:} SIXTE image and spectrum simulation with Athena/X-IFU for 50 ks. The spectrum in blue is extracted from a 4$\times$4 X-IFU pixels region (small green square in the Athena image). 
The XRISM/Resolve and Athena/X-IFU spectra have an adaptive rebinning to have at least 5 photons in each energy bin. We compare them with Chandra spectra (observations of 2009, 734 ks in total) extracted from the respective green regions.}
\label{fig:SimuSIXTE}
\end{figure*}

The spatially resolved spectroscopic study presented in the previous sections will be very useful to prepare observations at high spectral resolution in the X-ray band with XRISM (X-Ray Imaging and Spectroscopy Mission, launched in 2023) and Athena (Advanced Telescope for High ENergy Astrophysics, scheduled for 2037). This new generation of telescopes opens the door to high-resolution spectroscopy XRISM/Resolve and Athena/X-IFU (compared to $\sim$100 eV for Chandra/ACIS). Both instruments have spectral resolution under 5 eV but differ in number of pixels and telescope performances (effective area and  spatial resolution).

We used the SIXTE software package \citep{Dauser2019} provided by ECAP/Remeis observatory \footnote{\url{https://github.com/thdauser/sixte}} to simulate observations of Tycho's SNR as seen by XRISM/Resolve and Athena/X-IFU. 
For this high-resolution spectroscopy simulation, the observed line broadening modeled by the \textit{gsmooth} component (sufficient for the Chandra lower spectral resolution) had to be replaced by the contributions from the front and rear halves of the remnant shell.
We assign each half a model whose line-of-sight velocity is derived from the maps of averaged line-of-sight velocity captured by $z_{avg}$ and line broadening Sig$_{\rm 6keV}$ (shown in Fig. \ref{fig:Dynamics}) assuming the following relation : $z_{\rm front,\,rear} = z_{avg} \pm$ Sig$_{\rm 6keV}$/6.
Each front and rear models share the same plasma characteristics (temperature, ionization time, abundances, …) derived from our parameter maps but with their normalization halved.
With the tool \emph{simputfile} we simulated a Simput object with the aforementioned model for each of our 211 regions. 
After merging all these Simput objects, we simulated XRISM/Resolve and Athena/X-IFU observations with the tool \emph{xifupipeline}.
Fig \ref{fig:SimuSIXTE} presents simulations for the center of Tycho's SNR for 150 ks of XRISM/Resolve and 50 ks of Athena/X-IFU. The spectra are extracted from a region of 4$\times$4 pixels respectively for each instrument and are compared with Chandra spectra extracted in the same regions.

These simulations reveal that what we call lines in Fig \ref{fig:ExSpectre} will appear as multiple lines mainly associated with one element.
Some of the aims of XRISM will be to study the front and rear ejecta of the SNRs, thanks to the Doppler effect that will separate the two sides. It will be also possible to measure the $V_{\rm z}$ velocity of different elements. These individual lines will also allow to study the temperature of ions (in this article we measure the temperatures of electrons) by measuring the broadening of these lines \citep{Williams2020}.

However, the spatial resolution of XRISM will not be as good as Chandra's, which will bring new challenges to the X-ray analysis, such as the spatial-spectral mixing (contamination of the spectra from neighboring regions).
About four of our regions can be included in its Point Spread Function (PSF), so the mixing of zones with various physical properties is expected. For example, a broadening of the lines will happen due to the variations of redshift, and the temperature of the ions will be very challenging to evaluate. In Fig \ref{fig:CartoVNEI}, the temperature of the IMEs is more or less homogeneous in the center but the ionization time has some variability, which could change also the line ratios when averaging with the PSF.
This spatial-spectral mixing is expected to be less important for Athena/XIFU as shown in the simulated image in Fig \ref{fig:SimuSIXTE}.

The mapping presented in this study will help to evaluate these side effects and give keys to understanding the line forest that will be observed. This overview of the simulations that we can produce shows the incredible potential of XRISM/Resolve and Athena/X-IFU. More realistic simulations are beyond the scope of this paper as more hypotheses are needed.

\section{Summary}
\label{Section:Conclusion}

This study proposes a detailed parameter mapping of Tycho's SNR to investigate its origin.
The tessellation of the SNR in 211 regions, from which we extract the spectra, is based on the map of velocity along the line of sight $V_{\rm z}$ obtained in \cite{Godinaud2023}. Each region is then relatively homogeneous in velocity. 

Tycho's SNR contains a large variety of physical conditions across different regions and finding a unique spectral model to fit them is very challenging. We use an absorbed model with a thermal component for the IMEs, a hotter one for the iron-rich ejecta, a power law representing the synchrotron emission, and some additional Gaussians to improve the residuals. A spectral Gaussian smoothing is applied on the thermal components to mitigate the line broadening (mostly Si and S) due to the opposite dynamics on the two sides of the SNR.

The traditional methods of fitting like XSPEC showed their limits in front of the high dimensionality of our problem (19 degrees of freedom here). The results were not stable with respect to the initialization point and the computational time exploded during the estimation of the uncertainties. We used instead a Bayesian approach based on the nested sampling algorithm. The Bayesian X-ray analysis framework (BXA) allowed us to obtain for the first time the full posterior distribution for all 211 regions and so the uncertainties for all parameters.
We obtained good spectral fits with few residuals and very good constraints on each parameter.
The resulting parameter maps, obtained for the first time for some parameters, contain a wealth of information and our main results are emphasized below.

\begin{itemize}
    \item The initial objective of this paper was to investigate the large-scale asymmetry of $V_{\rm z}$, where the north ejecta of Tycho's SNR are more blueshifted and the south dominantly redshifted. This asymmetry can be innate (from an anisotropic explosion) or acquired (due to the interaction of the ejecta with some large structures in the ambient medium). We do not observe a similar dipole in our maps of temperature and ionization time.Thus, we cannot conclude about the origin of this asymmetry : a dipole during the explosion or a tilted equatorial overdensity will produce a velocity asymmetry but nothing similar on the other physical properties. 
    
    \item We map for the first time the photon index of the power law associated with non-thermal emission across the interior of the whole SNR. Our values concord with the bibliography for some special features but we measure also this parameter for some very faint filaments in the center of the SNR.
    
    \item A banana-shaped anti-correlation between the temperature and ionization time was observed in several SNRs, different telescopes, and spectral models.
    Despite the fact that there exists an underlying degeneracy due to the limited spectral resolution between k$T$ and $\tau$, the anti-correlation that we observe in our analysis is statistically significant. In addition, different regions of the SNR correspond to different sectors of the anti-correlation. It seems to indicate a physical origin of this relation associated to the history of the shocked ejecta.
    
    \item We compare the abundances of Mg, S, Ar, and Ca (with respect to Si), obtained in this study with some nucleosynthesis models. The values are coherent with the yields expected for a thermonuclear supernova explosion. 
    To discriminate the models, it will be interesting to compare our maps of abundances with the spatial production of the elements during the explosion. In the case of the delayed-detonation model of \cite{Seitenzahl2013}, we suspect an explosion with fewer than one hundred ignition points that produces an an-isotropic distribution of elements.
    It could explain the differences between the spatial distributions of Mg and the other IMEs that we observe.
    
    \item Finally, we detect and map for the first time the oxygen emission with a Gaussian component at 0.654 keV. It reveals an enhancement in the northeast. This oxygen can be some ejecta heated by the reverse shock but coming from a different layer than the IMEs and iron group, hence their different morphology. Or it can be some surrounding medium heated by the forward shock, a feature whose presence in the Tycho's SNR has not yet found a consensus in the literature.

\end{itemize}

\begin{acknowledgements}
The authors thank Hans Moritz Guenther, Laurence David and the Chandra/ACIS team for their advices about the sacrificial charge problem.
The authors are also very grateful to Maximilian Lorenz, Christian Kirsch and the SIXTE support team for their help to use the SIXTE software to conduct spectrum simulations.
The research leading to these results has received funding from the European Union’s Horizon 2020 Programme under the AHEAD2020 project (grant agreement n. 871158). This work was supported by CNES, focused on methodology for X-ray analysis.
\end{acknowledgements}


\begin{thebibliography}{45}
\expandafter\ifx\csname natexlab\endcsname\relax\def\natexlab#1{#1}\fi

\bibitem[{{Ashton} {et~al.}(2022){Ashton}, {Bernstein}, {Buchner}, {Chen}, {Cs{\'a}nyi}, {Fowlie}, {Feroz}, {Griffiths}, {Handley}, {Habeck}, {Higson}, {Hobson}, {Lasenby}, {Parkinson}, {P{\'a}rtay}, {Pitkin}, {Schneider}, {Speagle}, {South}, {Veitch}, {Wacker}, {Wales}, \& {Yallup}}]{Ashton2022}
{Ashton}, G., {Bernstein}, N., {Buchner}, J., {et~al.} 2022, Nature Reviews Methods Primers, 2, 39

\bibitem[{{Badenes} {et~al.}(2006){Badenes}, {Borkowski}, {Hughes}, {Hwang}, \& {Bravo}}]{Badenes2006}
{Badenes}, C., {Borkowski}, K.~J., {Hughes}, J.~P., {Hwang}, U., \& {Bravo}, E. 2006, \apj, 645, 1373

\bibitem[{{Ballet}(2006)}]{Ballet2006}
{Ballet}, J. 2006, Advances in Space Research, 37, 1902

\bibitem[{{Buchner}(2021)}]{Buchner2021}
{Buchner}, J. 2021, The Journal of Open Source Software, 6, 3001

\bibitem[{{Buchner}(2023)}]{Buchner2023}
{Buchner}, J. 2023, Statistics Surveys, 17, 169

\bibitem[{{Buchner} {et~al.}(2014){Buchner}, {Georgakakis}, {Nandra}, {Hsu}, {Rangel}, {Brightman}, {Merloni}, {Salvato}, {Donley}, \& {Kocevski}}]{Buchner2014}
{Buchner}, J., {Georgakakis}, A., {Nandra}, K., {et~al.} 2014, \aap, 564, A125

\bibitem[{{Cassam-Chena{\"\i}} {et~al.}(2007){Cassam-Chena{\"\i}}, {Hughes}, {Ballet}, \& {Decourchelle}}]{Cassam-Chenai2007}
{Cassam-Chena{\"\i}}, G., {Hughes}, J.~P., {Ballet}, J., \& {Decourchelle}, A. 2007, \apj, 665, 315

\bibitem[{{Chiotellis} {et~al.}(2013){Chiotellis}, {Kosenko}, {Schure}, {Vink}, \& {Kaastra}}]{Chiotellis2013}
{Chiotellis}, A., {Kosenko}, D., {Schure}, K.~M., {Vink}, J., \& {Kaastra}, J.~S. 2013, \mnras, 435, 1659

\bibitem[{{Dauser} {et~al.}(2019){Dauser}, {Falkner}, {Lorenz}, {Kirsch}, {Peille}, {Cucchetti}, {Schmid}, {Brand}, {Oertel}, {Smith}, \& {Wilms}}]{Dauser2019}
{Dauser}, T., {Falkner}, S., {Lorenz}, M., {et~al.} 2019, \aap, 630, A66

\bibitem[{{Decourchelle} {et~al.}(2001){Decourchelle}, {Sauvageot}, {Audard}, {Aschenbach}, {Sembay}, {Rothenflug}, {Ballet}, {Stadlbauer}, \& {West}}]{Decourchelle2001}
{Decourchelle}, A., {Sauvageot}, J.~L., {Audard}, M., {et~al.} 2001, \aap, 365, L218

\bibitem[{{Dickel} {et~al.}(1991){Dickel}, {van Breugel}, \& {Strom}}]{Dickel1991}
{Dickel}, J.~R., {van Breugel}, W.~J.~M., \& {Strom}, R.~G. 1991, \aj, 101, 2151

\bibitem[{{Diehl} \& {Statler}(2006)}]{diehl06b}
{Diehl}, S. \& {Statler}, T.~S. 2006, \mnras, 368, 497

\bibitem[{{Ellien} {et~al.}(2023){Ellien}, {Greco}, \& {Vink}}]{Ellien2023}
{Ellien}, A., {Greco}, E., \& {Vink}, J. 2023, \apj, 951, 103

\bibitem[{{Ferrand} {et~al.}(2019){Ferrand}, {Warren}, {Ono}, {Nagataki}, {R{\"o}pke}, \& {Seitenzahl}}]{Ferrand2019}
{Ferrand}, G., {Warren}, D.~C., {Ono}, M., {et~al.} 2019, \apj, 877, 136

\bibitem[{{Fruscione} {et~al.}(2006){Fruscione}, {McDowell}, {Allen}, {Brickhouse}, {Burke}, {Davis}, {Durham}, {Elvis}, {Galle}, {Harris}, {Huenemoerder}, {Houck}, {Ishibashi}, {Karovska}, {Nicastro}, {Noble}, {Nowak}, {Primini}, {Siemiginowska}, {Smith}, \& {Wise}}]{Fruscione2006}
{Fruscione}, A., {McDowell}, J.~C., {Allen}, G.~E., {et~al.} 2006, in Society of Photo-Optical Instrumentation Engineers (SPIE) Conference Series, Vol. 6270, Society of Photo-Optical Instrumentation Engineers (SPIE) Conference Series, ed. D.~R. {Silva} \& R.~E. {Doxsey}, 62701V

\bibitem[{{Godinaud} {et~al.}(2023){Godinaud}, {Acero}, {Decourchelle}, \& {Ballet}}]{Godinaud2023}
{Godinaud}, L., {Acero}, F., {Decourchelle}, A., \& {Ballet}, J. 2023, \aap, 680, A80

\bibitem[{{Greco} {et~al.}(2020){Greco}, {Vink}, {Miceli}, {Orlando}, {Dom{\v{c}}ek}, {Zhou}, {Bocchino}, \& {Peres}}]{Greco2020}
{Greco}, E., {Vink}, J., {Miceli}, M., {et~al.} 2020, \aap, 638, A101

\bibitem[{{Hwang} {et~al.}(2002){Hwang}, {Decourchelle}, {Holt}, \& {Petre}}]{Hwang2002}
{Hwang}, U., {Decourchelle}, A., {Holt}, S.~S., \& {Petre}, R. 2002, \apj, 581, 1101

\bibitem[{{Hwang} \& {Gotthelf}(1997)}]{Hwang1997}
{Hwang}, U. \& {Gotthelf}, E.~V. 1997, \apj, 475, 665

\bibitem[{{Kasuga} {et~al.}(2021){Kasuga}, {Vink}, {Katsuda}, {Uchida}, {Bamba}, {Sato}, \& {Hughes}}]{Kasuga2021}
{Kasuga}, T., {Vink}, J., {Katsuda}, S., {et~al.} 2021, \apj, 915, 42

\bibitem[{{Kerzendorf} {et~al.}(2018){Kerzendorf}, {Long}, {Winkler}, \& {Do}}]{Kerzendorf2018}
{Kerzendorf}, W.~E., {Long}, K.~S., {Winkler}, P.~F., \& {Do}, T. 2018, \mnras, 479, 5696

\bibitem[{{Krause} {et~al.}(2008){Krause}, {Tanaka}, {Usuda}, {Hattori}, {Goto}, {Birkmann}, \& {Nomoto}}]{Krause2008}
{Krause}, O., {Tanaka}, M., {Usuda}, T., {et~al.} 2008, \nat, 456, 617

\bibitem[{{Li} {et~al.}(2015){Li}, {Decourchelle}, {Miceli}, {Vink}, \& {Bocchino}}]{Li2015}
{Li}, J.-T., {Decourchelle}, A., {Miceli}, M., {Vink}, J., \& {Bocchino}, F. 2015, \mnras, 453, 3953

\bibitem[{{Liu} {et~al.}(2023){Liu}, {R{\"o}pke}, \& {Han}}]{Liu2023}
{Liu}, Z.-W., {R{\"o}pke}, F.~K., \& {Han}, Z. 2023, Research in Astronomy and Astrophysics, 23, 082001

\bibitem[{{Lu} {et~al.}(2011){Lu}, {Wang}, {Ge}, {Qu}, {Yang}, {Zheng}, \& {Chen}}]{Lu2011}
{Lu}, F.~J., {Wang}, Q.~D., {Ge}, M.~Y., {et~al.} 2011, \apj, 732, 11

\bibitem[{{Matsuda} {et~al.}(2020){Matsuda}, {Tanaka}, {Uchida}, {Amano}, \& {Tsuru}}]{Matsuda2020}
{Matsuda}, M., {Tanaka}, T., {Uchida}, H., {Amano}, Y., \& {Tsuru}, T.~G. 2020, \pasj, 72, 85

\bibitem[{{Mayer} {et~al.}(2023){Mayer}, {Becker}, {Predehl}, \& {Sasaki}}]{Mayer2023}
{Mayer}, M. G.~F., {Becker}, W., {Predehl}, P., \& {Sasaki}, M. 2023, \aap, 676, A68

\bibitem[{{Millard} {et~al.}(2020){Millard}, {Bhalerao}, {Park}, {Sato}, {Hughes}, {Slane}, {Patnaude}, {Burrows}, \& {Badenes}}]{Millard2020}
{Millard}, M.~J., {Bhalerao}, J., {Park}, S., {et~al.} 2020, \apj, 893, 98

\bibitem[{{Millard} {et~al.}(2022){Millard}, {Park}, {Sato}, {Hughes}, {Slane}, {Patnaude}, {Burrows}, \& {Badenes}}]{Millard2022}
{Millard}, M.~J., {Park}, S., {Sato}, T., {et~al.} 2022, \apj, 937, 121

\bibitem[{{Okuno} {et~al.}(2020){Okuno}, {Tanaka}, {Uchida}, {Aharonian}, {Uchiyama}, {Tsuru}, \& {Matsuda}}]{Okuno2020}
{Okuno}, T., {Tanaka}, T., {Uchida}, H., {et~al.} 2020, \apj, 894, 50

\bibitem[{{Sanders}(2006)}]{sanders06}
{Sanders}, J.~S. 2006, \mnras, 371, 829

\bibitem[{{Sato} \& {Hughes}(2017)}]{Sato2017a}
{Sato}, T. \& {Hughes}, J.~P. 2017, \apj, 840, 112

\bibitem[{{Seitenzahl} {et~al.}(2013){Seitenzahl}, {Ciaraldi-Schoolmann}, {R{\"o}pke}, {Fink}, {Hillebrandt}, {Kromer}, {Pakmor}, {Ruiter}, {Sim}, \& {Taubenberger}}]{Seitenzahl2013}
{Seitenzahl}, I.~R., {Ciaraldi-Schoolmann}, F., {R{\"o}pke}, F.~K., {et~al.} 2013, \mnras, 429, 1156

\bibitem[{{Sun} \& {Chen}(2019)}]{Sun2019}
{Sun}, L. \& {Chen}, Y. 2019, \apj, 872, 45

\bibitem[{{Tran} {et~al.}(2015){Tran}, {Williams}, {Petre}, {Ressler}, \& {Reynolds}}]{Tran2015}
{Tran}, A., {Williams}, B.~J., {Petre}, R., {Ressler}, S.~M., \& {Reynolds}, S.~P. 2015, \apj, 812, 101

\bibitem[{{Uchida} {et~al.}(2024){Uchida}, {Kasuga}, {Maeda}, {Lee}, {Tanaka}, \& {Bamba}}]{Uchida2024}
{Uchida}, H., {Kasuga}, T., {Maeda}, K., {et~al.} 2024, \apj, 962, 159

\bibitem[{{Williams} {et~al.}(2013){Williams}, {Borkowski}, {Ghavamian}, {Hewitt}, {Mao}, {Petre}, {Reynolds}, \& {Blondin}}]{Williams2013}
{Williams}, B.~J., {Borkowski}, K.~J., {Ghavamian}, P., {et~al.} 2013, \apj, 770, 129

\bibitem[{{Williams} {et~al.}(2016){Williams}, {Chomiuk}, {Hewitt}, {Blondin}, {Borkowski}, {Ghavamian}, {Petre}, \& {Reynolds}}]{Williams2016}
{Williams}, B.~J., {Chomiuk}, L., {Hewitt}, J.~W., {et~al.} 2016, \apjl, 823, L32

\bibitem[{{Williams} {et~al.}(2017){Williams}, {Coyle}, {Yamaguchi}, {Depasquale}, {Seitenzahl}, {Hewitt}, {Blondin}, {Borkowski}, {Ghavamian}, {Petre}, \& {Reynolds}}]{Williams2017}
{Williams}, B.~J., {Coyle}, N.~M., {Yamaguchi}, H., {et~al.} 2017, \apj, 842, 28

\bibitem[{{Williams} {et~al.}(2020){Williams}, {Katsuda}, {Cumbee}, {Petre}, {Raymond}, \& {Uchida}}]{Williams2020}
{Williams}, B.~J., {Katsuda}, S., {Cumbee}, R., {et~al.} 2020, \apjl, 898, L51

\bibitem[{{Wilms} {et~al.}(2000){Wilms}, {Allen}, \& {McCray}}]{Wilms2000}
{Wilms}, J., {Allen}, A., \& {McCray}, R. 2000, \apj, 542, 914

\bibitem[{{Yamaguchi} {et~al.}(2017){Yamaguchi}, {Hughes}, {Badenes}, {Bravo}, {Seitenzahl}, {Mart{\'\i}nez-Rodr{\'\i}guez}, {Park}, \& {Petre}}]{Yamaguchi2017}
{Yamaguchi}, H., {Hughes}, J.~P., {Badenes}, C., {et~al.} 2017, \apj, 834, 124

\bibitem[{{Yamaguchi} {et~al.}(2011){Yamaguchi}, {Koyama}, \& {Uchida}}]{Yamaguchi2011}
{Yamaguchi}, H., {Koyama}, K., \& {Uchida}, H. 2011, \pasj, 63, S837

\bibitem[{{Zhou} {et~al.}(2016){Zhou}, {Chen}, {Zhang}, {Li}, {Safi-Harb}, {Zhou}, \& {Zhang}}]{Zhou2016}
{Zhou}, P., {Chen}, Y., {Zhang}, Z.-Y., {et~al.} 2016, \apj, 826, 34

\bibitem[{{Zhou} \& {Vink}(2018)}]{Zhou2018}
{Zhou}, P. \& {Vink}, J. 2018, \aap, 615, A150

\end{thebibliography}

\begin{appendix} 

\section{Segmentation}
\label{Appendix : Segmentation}

\begin{figure*}[h]
\begin{subfigure}{0.5\textwidth}
\includegraphics[scale=0.24, trim = 0 0 0 0, clip=true]{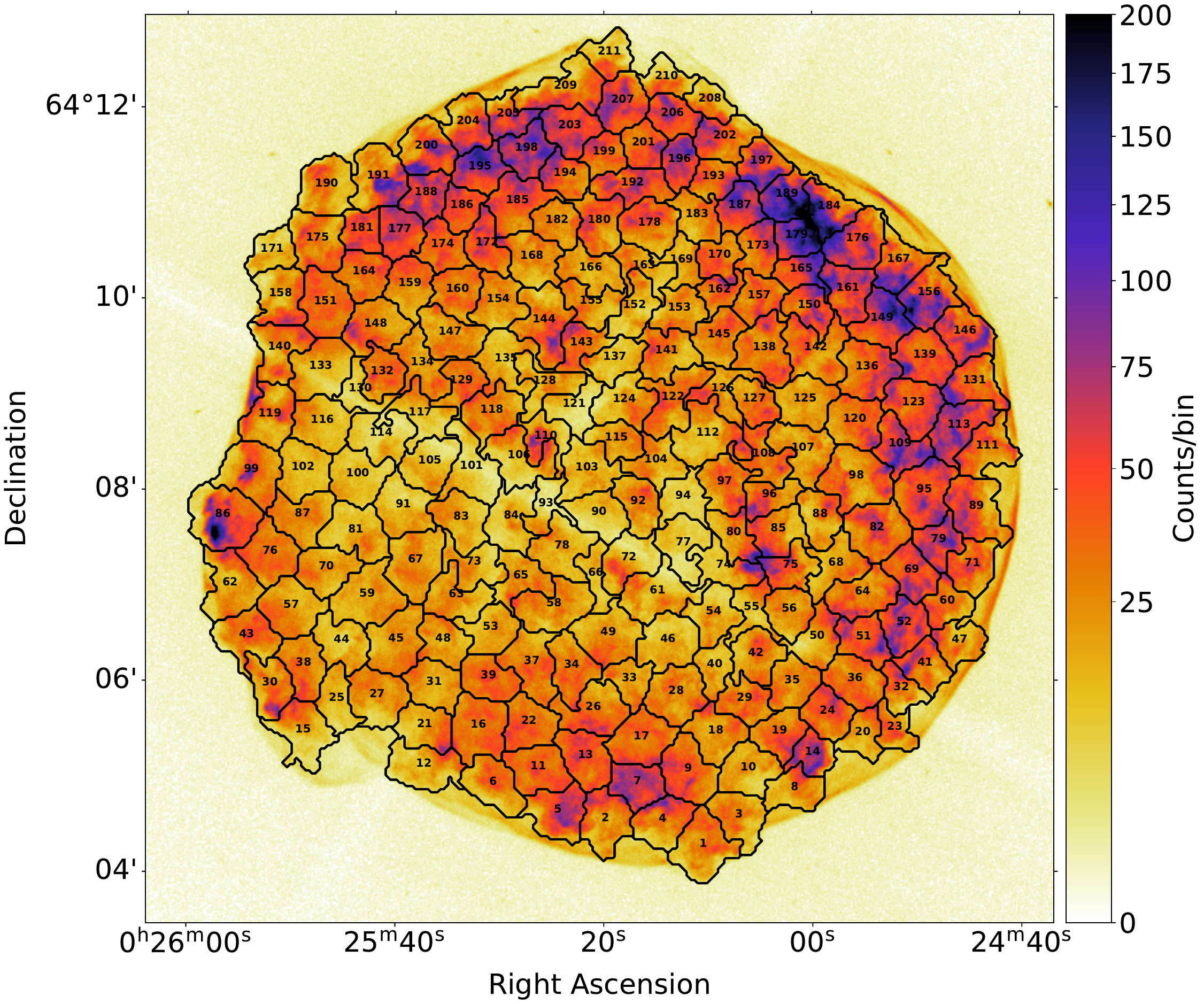} 
\end{subfigure}
\begin{subfigure}{0.5\textwidth}
\includegraphics[scale=0.24, trim = 0 0 0 0, clip=true]{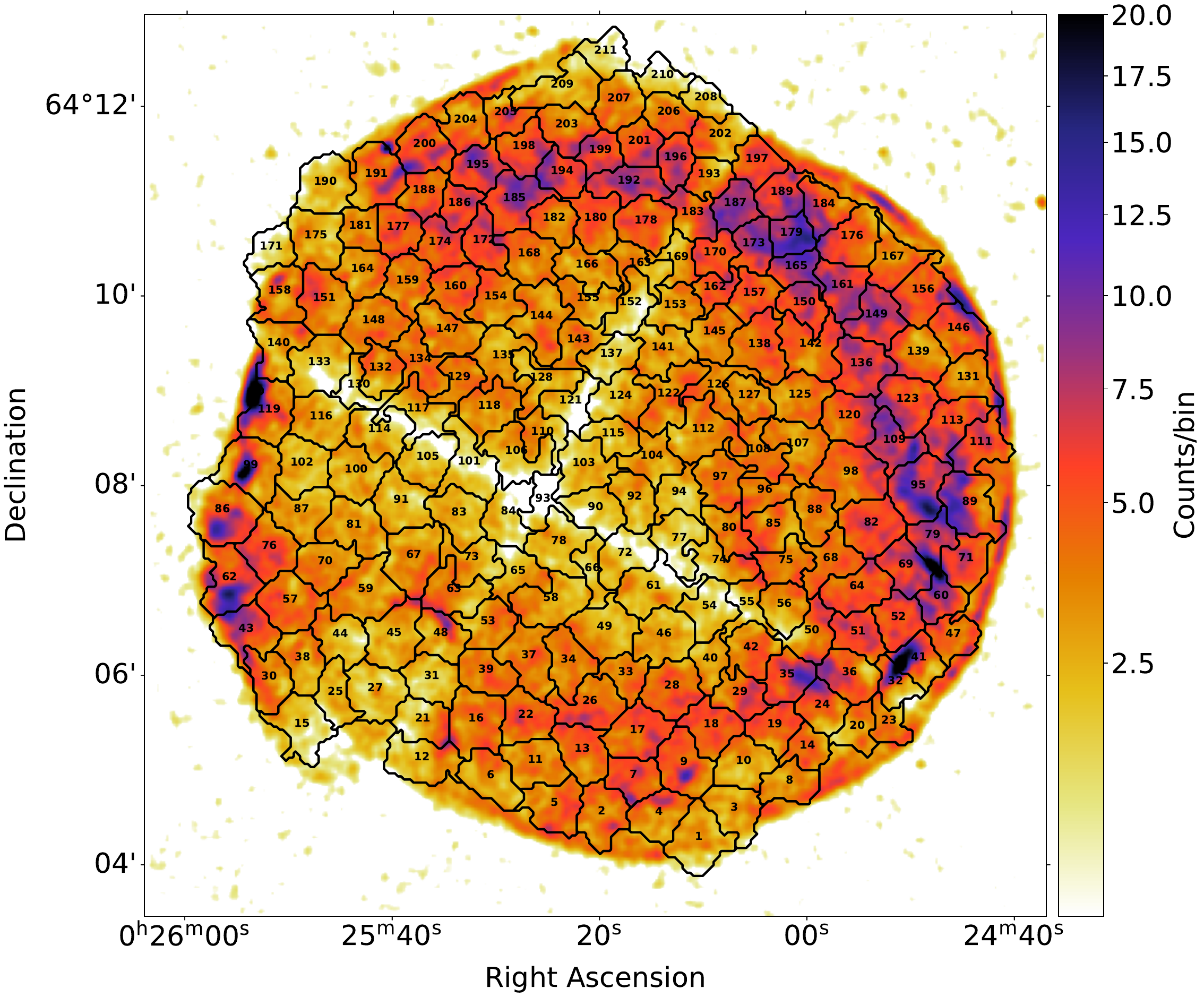}
\end{subfigure}
\caption{\footnotesize 
The segmentation used in this study compared to the image of Tycho's SNR in different energy bands. The map is not corrected from the exposure map to highlight the diversity in counts across the regions. \emph{Left : } Between 1.6 and 4.2 keV to highlight the IME emission (Si, S, Ar and Ca) with a bin size of 0.5" \emph{Right : } An energy band between 5.8 and 7.1 keV, which represents the Fe-K emission with some contribution of the synchrotron emission with a bin size of 2" and a Gaussian smoothing with a kernel of 2".}
\label{fig:App1}
\end{figure*}

For further information, in Fig \ref{fig:App1}, we compare our tesselation of 211 regions with maps of the flux of Tycho's SNR in two energy bands : [1.6, 4.2] keV dominated by IME emission and [5.8, 7.1] keV for the Fe-K line. These maps allow to check if there are bright knots in a region. They could have different properties from the global ejecta emission and explain an anomaly in comparison to the surrounding regions.

\section{Examples of posterior distribution}
\label{Appendix:CornerPlot}

\begin{figure*}
\centering
\includegraphics[scale=0.34, trim = 0 0 0 0, clip=true]{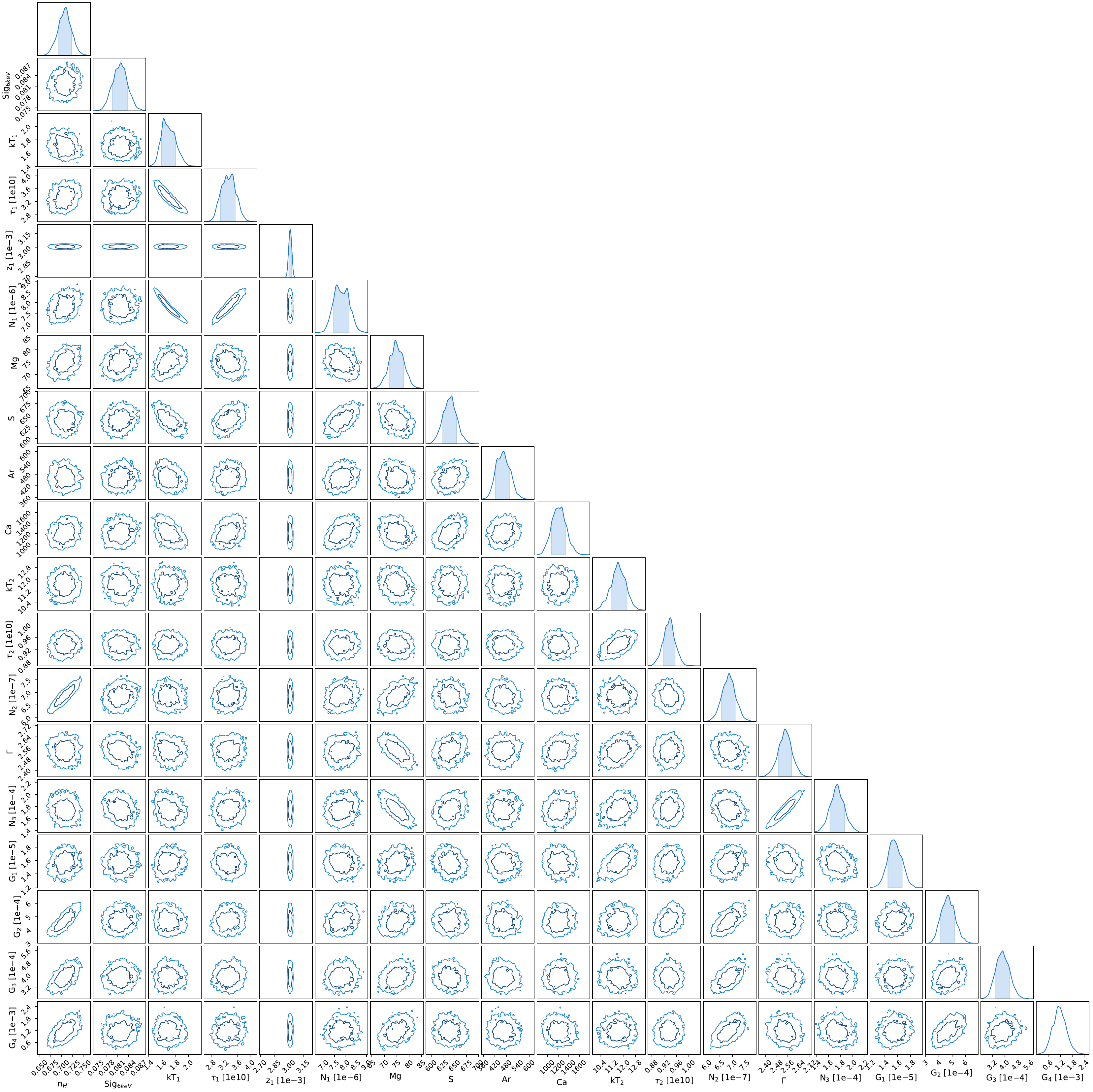}
\caption{\footnotesize The corner plot of the region 59, showing the posterior distributions obtained with BXA for all the parameters of our spectral model. The corresponding spectrum is shown in Fig \ref{fig:ExSpectre}. The contours corresponds to 68\% and 95\% confidence levels.}
\label{fig:CornerPlot59}
\end{figure*}

\begin{figure*}[h]
\includegraphics[scale=0.4, trim = 0 0 0 0, clip=true]{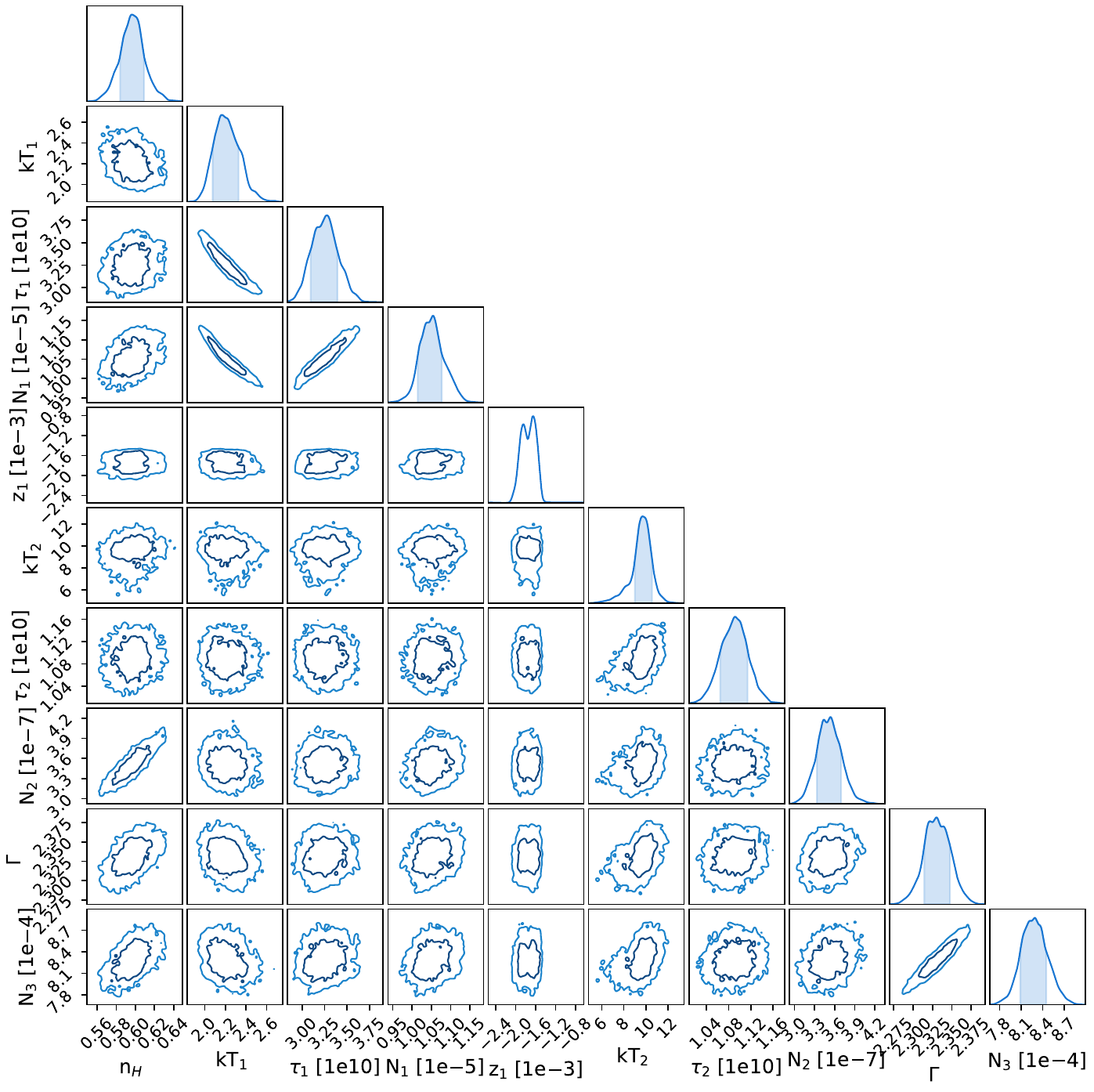} 
\includegraphics[scale=0.4, trim = 0 0 0 0, clip=true]{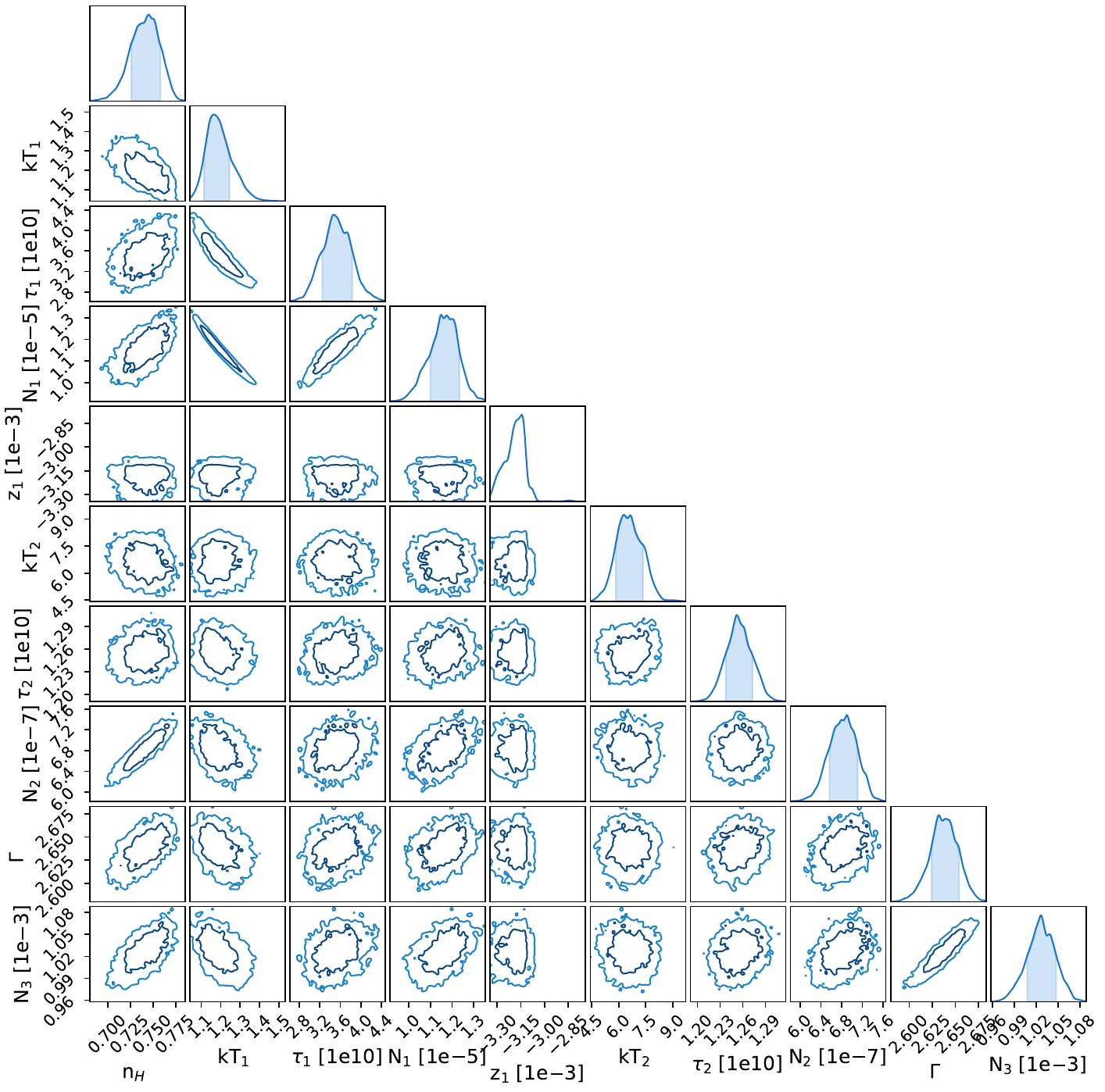}
\includegraphics[scale=0.4, trim = 0 0 0 0, clip=true]{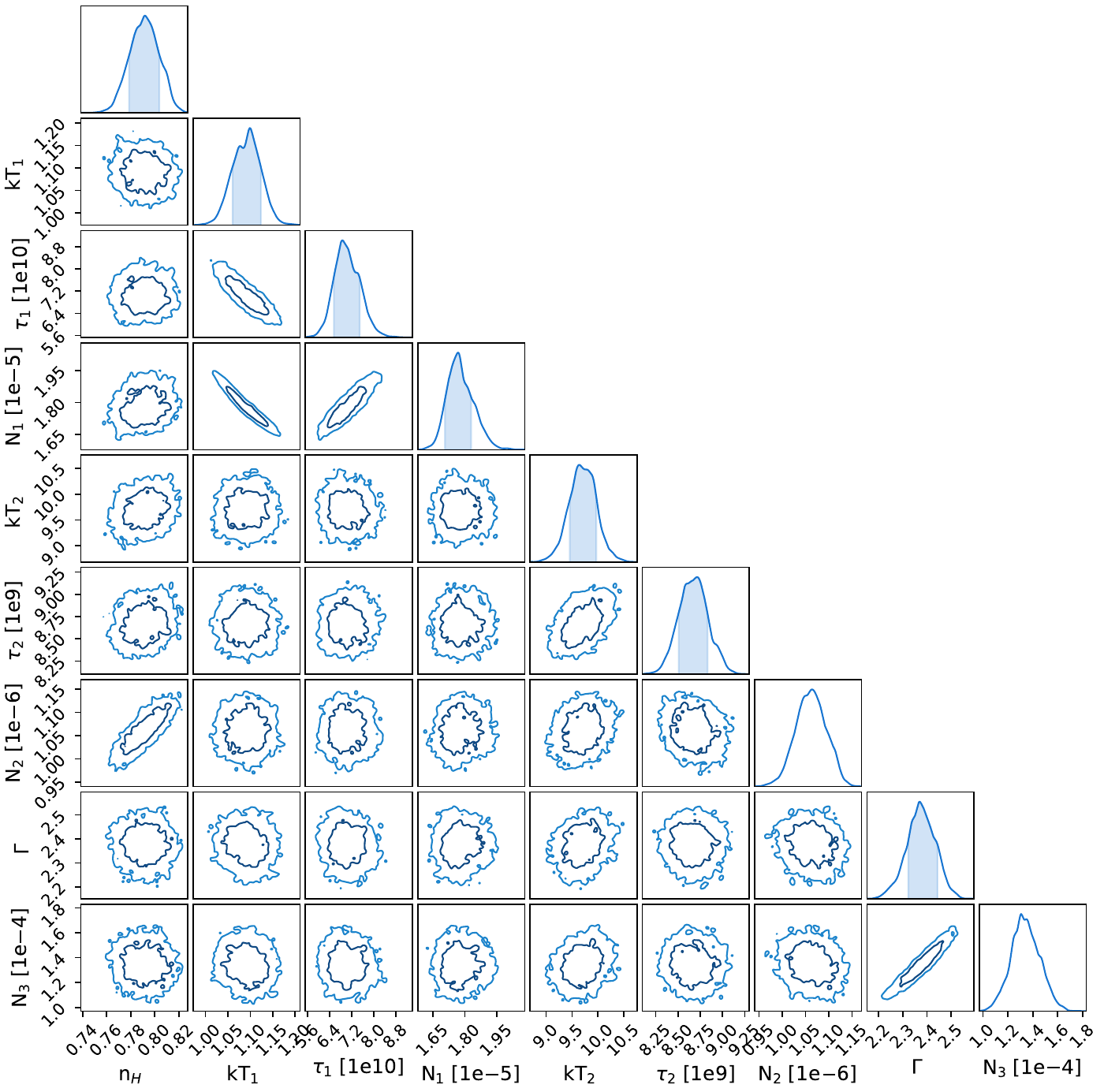} 
\includegraphics[scale=0.4, trim = 0 0 0 0, clip=true]{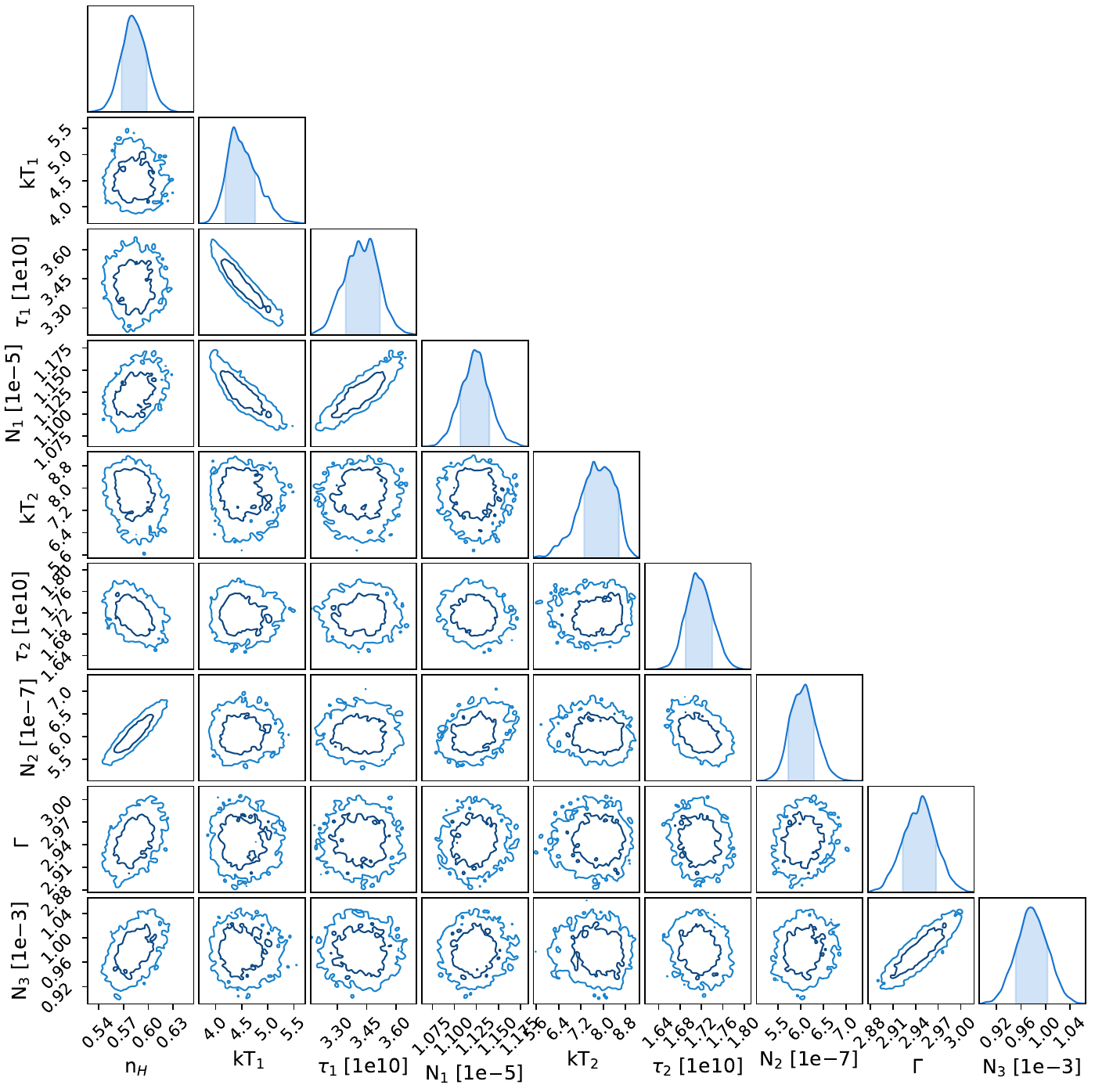}

\caption{\footnotesize Four other examples of posterior distribution of the regions 89 \emph{(top left)}, 99 \emph{(top right)}, 174 \emph{(bottom left)} and 184 \emph{(bottom right)} whose spectra are shown in Fig \ref{fig:ExSpectre}. The number of parameters is reduced here only for visualization, the posterior projections not shown here are also well constrained.} 
\label{fig:CornerPlot89_99}
\end{figure*}

We present in this appendix examples of the posterior distribution obtained with the tool BXA (see Sect \ref{Subsection:BXA}). The spectra associated with these regions are shown in Fig \ref{fig:ExSpectre}. We can see in Fig \ref{fig:CornerPlot59} the complete posterior distribution for region 59. This corner plot shows that each parameter is well-constrained. It was a true challenge to check by eye all this information for the 211 regions. A challenge also of visualization in an article, in Fig \ref{fig:CornerPlot89_99}, we choose to present four other examples (regions 89, 99, 174 and 184) but reduced to only the main parameters.

In these five examples, we observe some notable correlations: within the group k$T_1$ - $\tau_1$ - norm of the IME component (N$_1$ in the corner plot), between $N_{\rm H}$ and the norm of the iron-rich VNEI (N$_2$ in the corner plot) and between the photon index ($\Gamma$) and the norm of the power law (N$_3$ in the corner plot). These are expected because the same part of the spectrum is used to constrain these parameters. However, we can see that uncertainties are small as we have a lot of statistics in each region.

\section{Maps of relative errors}
\label{Appendix:ErreurRelative}

\begin{figure*}
\centering
\includegraphics[scale=0.6, trim = 0 0 0 0, clip=true]{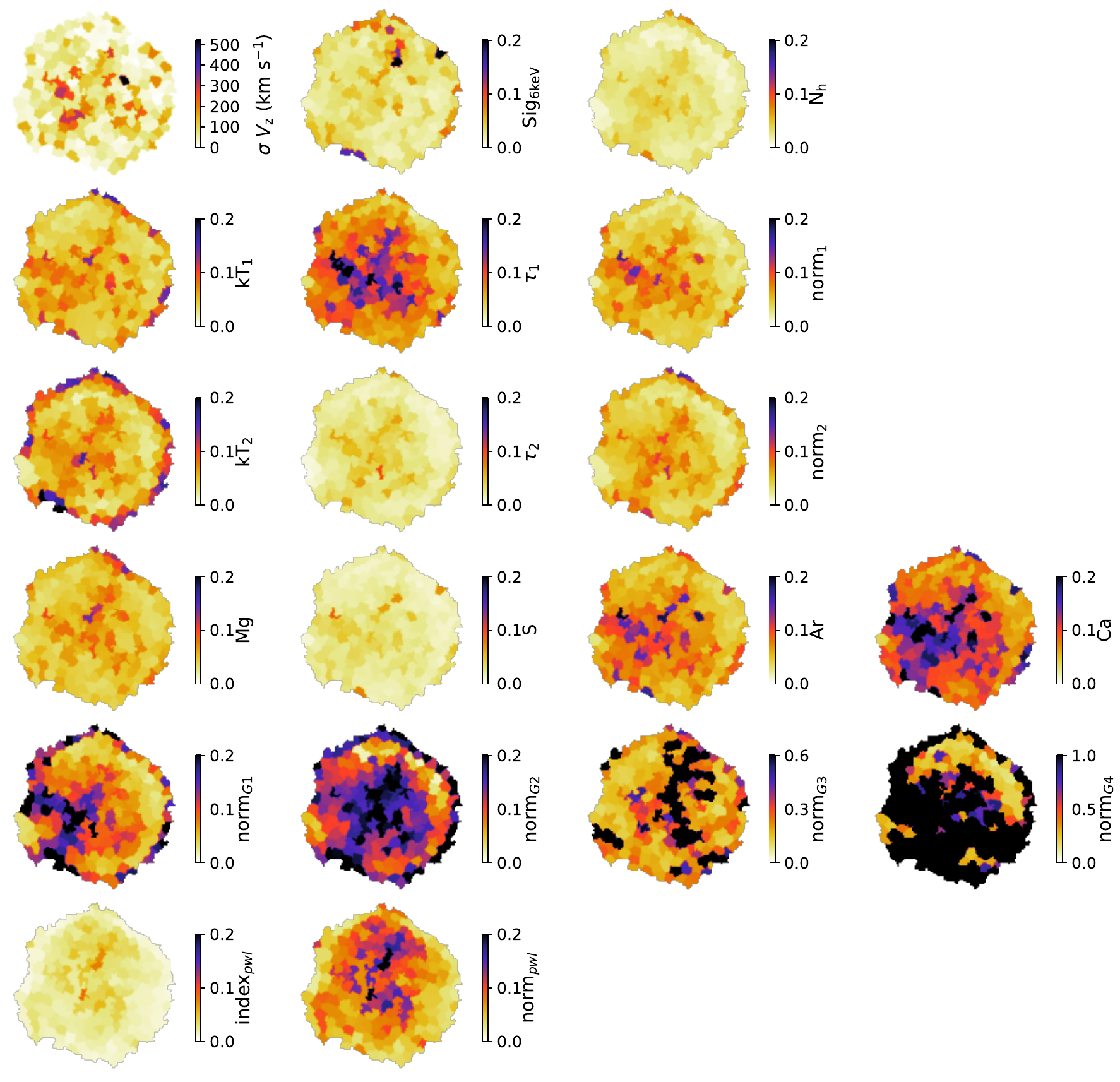}
\caption{\footnotesize Maps of relative errors for the 19 parameters fitted in this study. The error maps in this figure follow the order of the parameter maps in this article. Note that for the velocity, we plot the absolute error (more relevant) instead. The norms of G3 and G4 do not share the same color scaling as the other parameters. }
\label{fig:ErreurRelative}
\end{figure*}

To have a more precise view of the uncertainties for each parameter, we map in Fig \ref{fig:ErreurRelative} the relative error in each region, defined as the standard deviation of the posterior distribution divided by its median. For the redshift map, we plot the standard deviation map of $V_{\rm z}$, as the absolute error is the relevant quantity. 
Most of the parameters have a relative error less than 10\%. Nevertheless the norm of the additional Gaussians, in particular G3 and G4 are poorly constrained in a lot of regions as observed also in Fig \ref{fig:ExSpectre}.

The regions of scientific interest highlighted in this study are uncorrelated with the features visible in these error maps. In general, the regions with bigger uncertainties are located in the gap between the CCDs of Chandra and in the southeast of Tycho's SNR, where the statistics are lower.

\end{appendix}

\end{document}